\documentclass[10pt,conference]{IEEEtran}
\IEEEoverridecommandlockouts
\usepackage{framed}
\usepackage{listings}
\usepackage{subcaption}
\usepackage{longtable}
\usepackage{microtype}
\usepackage{lipsum}
\usepackage{comment}
\usepackage{mdframed}
\usepackage{colortbl}
\usepackage{collcell}
\usepackage{adjustbox}
\usepackage[ruled,vlined,linesnumbered]{algorithm2e}
\usepackage{amsfonts}
\usepackage{mathtools}
\usepackage{braket}
\usepackage{mathpartir}
\usepackage{bm}
\usepackage[dvipsnames]{xcolor}
\usepackage{xparse}
\usepackage{etoolbox}
\usepackage{tabularx}
\usepackage{xfrac}
\usepackage{booktabs}
\usepackage{xspace}
\usepackage{tikz-cd}
\usepackage{pgf}
\usepackage{tikz}
\usepackage{wrapfig}
\usepackage{subcaption}
\usepackage{pifont}
\usepackage{hhline}
\usepackage{multirow}
\usepackage{enumitem}
\usepackage{ragged2e}
\usepackage{fancyvrb}
\DeclareMathOperator{\dplus}{+\kern -0.4em+}

\DeclareUnicodeCharacter{21D2}{$~\Rightarrow$}
\DeclareUnicodeCharacter{202F}{}

\usepackage[colorlinks=true, urlcolor=blue, linkcolor=red]{hyperref}

\usepackage{commands}

\begin{document}

\title{LLM-Assisted Synthesis of High-Assurance C Programs}

\author{
  \IEEEauthorblockN{Prasita Mukherjee}
  \IEEEauthorblockA{\textit{Department of Computer Science} \\
  \textit{Purdue University}\\
    West Lafayette, USA \\
    mukher39@purdue.edu}
\and
\IEEEauthorblockN{Minghai Lu}
  \IEEEauthorblockA{\textit{Department of Computer Science} \\
  \textit{Purdue University}\\
    West Lafayette, USA \\
    lu1074@purdue.edu}
\and
  \IEEEauthorblockN{Benjamin Delaware}
  \IEEEauthorblockA{\textit{Department of Computer Science} \\
  \textit{Purdue University}\\
    West Lafayette, USA \\
    bendy@purdue.edu}
}
\maketitle

\begin{abstract}
  We present \synver{} --- a novel, general purpose synthesizer for C
  programs equipped with machine-checked proofs of correctness using
  the Verified Software Toolchain. To do so, \synver{} employs two
  Large Language Models (LLMs): the first generates candidate programs
  from user-provided specifications, and the second helps
  automatically construct formal proofs of their correctness in the
  Rocq proof assistant. To facilitate verification, \synver{} places a
  set of syntactic restrictions on candidate programs that make them
  amenable to automated reasoning. \synver{} uses a hybrid
  verification strategy that combines symbolic reasoning with
  LLM-powered proof generation to discharge proof obligations that the
  symbolic engine cannot handle on its own. We demonstrate the
  applicability of \synver{} using a diverse set of benchmarks drawn
  from the program synthesis and verification literature.
\end{abstract}


\section{Introduction}
\label{sec:intro}
The goal of \emph{program synthesis} is to automatically generate a
program from a high-level specification of its intended
behavior~\cite{MW+79}. While the form of these specifications can
vary, e.g., input-output examples that describe a subset of the target
program's functionality, or a logical formula that fully captures the
desired behavior, traditional program synthesis techniques typically
guarantee that generated programs satisfy the input
specification. \emph{Deductive synthesis} techniques in particular aim
to provide strong guarantees by framing synthesis as a deductive
inference problem, and employ a rule-based search to find a program
that meets the target specification. These systems often repurpose
program \emph{verification} rules to ensure that each step is
justified, resulting in programs that are \textit{correct by
  construction}. Deductive techniques have been successfully applied
to a diverse set of domains, including SQL-style
queries~\cite{fiatPOPL2015}, heap-manipulating
programs~\cite{suslik2019, suslikICFP21}, serializers and
deserializers~\cite{narcissusDelaware2019}, clients of APIs with
strong specifications~\cite{cobaltMishra22}, and concurrent garbage
collectors~\cite{SpecWareGC}. These rigorous guarantees come at a
cost, however: to keep this search tractable, fully automated tools
are forced to limit the class of specifications and programs they can
handle.

More recently, large language models (LLMs) have shown a remarkable
ability to automatically generate programs from natural language
descriptions, and have quickly become part of the modern software
development toolbox. Unlike traditional program synthesis techniques,
however, LLM-powered code generation tools do not provide any
guarantees about the behaviors of the programs they produce, leaving
that task entirely to the user. In response, several recent works have
investigated how to provide more assurance about LLM-generated
programs, e.g., by targeting verification-oriented languages like
Dafny~\cite{LLMDafny2024} and F*~\cite{chakraborty2024towards}. In the
case of tools that target mainstream languages such as C and Rust,
several works have proposed adding annotations to programs that can
then be statically checked by existing automated
verifiers~\cite{AutoVerus,veCoGen,autoSpecCAV24}. Unfortunately, these
annotation-based approaches target assertion logics that are not
expressive enough to capture the full range of specifications used by
prior deductive synthesizers, including properties of heap
manipulating programs expressed in separation
logic~\cite{reynolds2002}.

This work proposes \synver{}
, a framework that fills this gap by
combining LLM-powered code and proof generation with deductive
verification. \synver{} synthesizes C programs from semantically rich
specifications, with machine-checked guarantees about their
correctness. Unlike prior deductive synthesis engines, which are
typically domain-specific, our tool is expressive enough to support
multiple application domains. 
To do so, we leverage two key insights from prior deductive synthesis
approaches: first, we constrain the space of candidate solutions by
automated verification. Second, we develop custom proof automation
procedures, or \emph{tactics}, tailored to programs meeting this bias,
enabling automated verification using an existing (interactive)
verification framework~\cite{vst2018appel} implemented in the Rocq/Coq
proof assistant~\cite{Coq-refman}. Due to the richness of our
specification language, our tactic-based automation is necessarily
incomplete; we address this limitation by introducing a novel
LLM-powered proof synthesis technique~\cite{yang2023leandojo,
  baldur23} that discharges proof obligations that our tactic cannot
resolve on its own.

We evaluate \synver{} on three synthesis problems drawn from three
distinct application domains previously targeted by separate deductive
synthesis engines. While each of those tools can only handle problems
from their particular domain, our tool is flexible enough to
synthesize high-assurance C programs for all three domains. Our
experiments also show that our combination of custom proof automation
and LLM-based proof synthesis outperforms existing LLM-based proof
automation techniques when reasoning about the correctness of programs
generated by \synver{}.

In summary, this paper presents the following contributions:
\begin{itemize}
\item We propose a novel LLM-powered program synthesis framework that
  generates high-assurance C programs from rich logical
  specifications, with machine-checked proofs of their correctness in
  the Rocq proof assistant.
\item We show that by biasing the space of candidate programs and
  combining custom proof automation and LLM-based proof synthesis, we
  can repurpose an existing interactive program verification framework
  to automatically verify LLM-generated programs.
\item We demonstrate the flexibility of our framework by evaluating it
  on a suite of benchmarks drawn from three distinct domains from the
  deductive synthesis literature, and show that our hybrid proof
  automation approach outperforms prior LLM-based proof techniques
  when reasoning about these programs. An artifact containing the
  source code of \synver{} and our experiments is publicly
  available~\cite{synver:artifact}.
\end{itemize}



\section{Background}
\label{sec:overview}
We begin by briefly reviewing the Verification Software Toolchain
(VST)~\cite{vst2018appel}, the Rocq/Coq-based program verification
framework that \synver{} uses to reason about the C programs it
generates. 
In VST, program properties are expressed using Hoare
triples~\cite{hoare1969} of the form $\vdash \{P\} \;$ \cinline{c}
$\; \{Q\}$ which claims that when the program \cinline{c} is executed
in a state satisfying the precondition $P$, it will either run forever
or terminate in a state satisfying the postcondition $Q$. VST is
equipped with a separation logic for proving that such triples are
valid. It also includes a set of \emph{tactics}~\cite{cao2018vst} that
developers can use to interactively build up proofs of program
correctness using the rules of the underlying separation logic.



\begin{figure}[!t]
  \begin{center}
    \begin{cprog}
  /* {h1 |$\mapsto$| l1 * h2 |$\mapsto$| l2} */
  struct sll* append(struct sll* h1, struct sll* h2)
  {
    struct sll *current = h1;
    if (h1 == NULL) {
        return h2;
    }
    while (current->next != NULL) {
        current = current->next;
    }
    current->next = h2;
    return h1;
  }
  /* {h |$\mapsto$| (l1 |$\dplus$| l2)} /*
\end{cprog}
\end{center}
\caption{C function that concatenates two singly-linked lists}
\vspace{-1em}
\label{fig:app+Impl}
\end{figure}

To illustrate this process, consider the \cinline|append| function
shown in \autoref{fig:app+Impl}. When given pointers to two valid
singly linked lists, \cinline|append| concatenates them together and
returns a pointer to the head of the resulting (singly linked)
list. The comments on lines 1 and 14 give pre- and post-conditions
that specify this behavior. To verify that \cinline|append| meets this
specification using VST, a user first defines a \textit{theorem}
stating a Hoare triple with these pre- ($P$) and postconditions ($Q$),
as shown on lines 2-3 of \autoref{fig:loopInvariant}. Processing this
definition causes Rocq to enter its interactive proof mode, displaying
an initial \emph{goal}, or proof obligation--- in this case, the
top-level correctness statement for \cinline|append|. The user then
writes a \textit{proof script}, a sequence of tactics that explain how
to build a proof of this goal; lines 5-19 of
\autoref{fig:loopInvariant} show the first part of a proof script for
\texttt{append\_correct}. Processing a tactic replaces the current
goal with a (possibly empty) set of new subgoals, again shown to the
user; the proof is complete when no subgoals remain.

The proof script in \autoref{fig:loopInvariant} includes both
VST-specific and Rocq's built-in tactics, which are highlighted in
\textcolor{purple}{purple} and \textcolor{ForestGreen}{green},
respectively.  Some tactics take arguments: \rocqinline{rewrite} on
line 9, for example, takes a fact of the form \rocqinline|x = y|, and
replaces all occurrences of \rocqinline|x| in the current goal with
\rocqinline|y|. VST-provided tactics often require richer inputs: the
\rocqinline{|\textcolor{purple}{\textbf{forward\_loop}}|} tactic on
line 11, for example, takes an inductive \emph{loop invariant} that
specifies the behavior of each iteration of the loop on lines 8-10 of
\cinline|append|. Automatically coming up with loop invariants is one
of the most challenging problems in program
verification~\cite{furia2014loop}, and VST thus expects users to
supply them manually. Tactics can fail if applied to goals that do not
have the expected shape. Oftentimes, users will apply tactics to
change a goal into one that a specific tactic can handle: the tactics
on lines 5-7, for instance, transform the goal into a form supported
by \rocqinline{entailer!!}. The tactics on line 9 similarly produce a
goal which \rocqinline{forward} can process.


\begin{figure}[!t]
  \centering
  \begin{rocq}
Theorem append_correct :
  |$\vdash$| - {SEP(listrep l1 h1; listrep l2 h2)} append
  {Exists h: val, SEP(listrep (l1 |$\dplus$| l2) h)}.

|\textcolor{purple}{\textbf{forward\_if}}|. |\textcolor{purple}{\textbf{forward}}|.
assert (11 = nil Z). apply H0. reflexivity.
subst. |\textcolor{purple}{\textbf{Exists}}| h2. |
\textcolor{purple}{\textbf{entailer!!}}|. simpl. |\textcolor{purple}{\textbf{entailer!!}}|
rewrite (listrep_nonnull_h1) by auto. |\textcolor{purple}{\textbf{Intros}}| h hs y. |
\textcolor{purple}{\textbf{forward}}|.
|\textcolor{purple}{\textbf{forward\_loop}}|
  (EX sla: list Z, EX b: Z, EX sic: list Z,
   EX t: val, EX u: val,
   PROP (Int.min_signed <= b <= Int.max_signed;
             11 = sla ++ b :: s1c)
   LOCAL (temp _h1 h1; temp_h2 h2; temp_head t)
   SEP (lseg sla pl t;
          data_at Ish t_list (Vint (Int. repr b), u) t;
          listrep sic u; listrep 12 h2))
  \end{rocq}

  \caption{A partial proof of correctness of \lstinline|append| in
    VST}
  \label{fig:loopInvariant}
  \vspace{-1em}
\end{figure}

As this example illustrates, VST is a powerful tool for constructing a
formal proof about rich behaviors of C programs. The framework is
designed to be used interactively, with a user inspecting the current
goal and supplying a tactic that moves the proof forward, e.g., by
providing loop invariants or transforming goals into a form that
VST-supplied tactics can automatically discharge. \synver{} builds on
top of the foundation provided by VST to verify generated programs,
but attempts to remove the user from the loop.
The next section describes how \synver{} constrains the shape of generated
programs to make them amenable to automated verification, uses custom
proof automation to discharge many of the resulting proof obligations,
and deploys LLM-guided proof synthesis to handle the rest.


\section{Approach}
\label{sec:approach}

\begin{figure}[!htb]
  \centering
  \includegraphics[scale=.25]{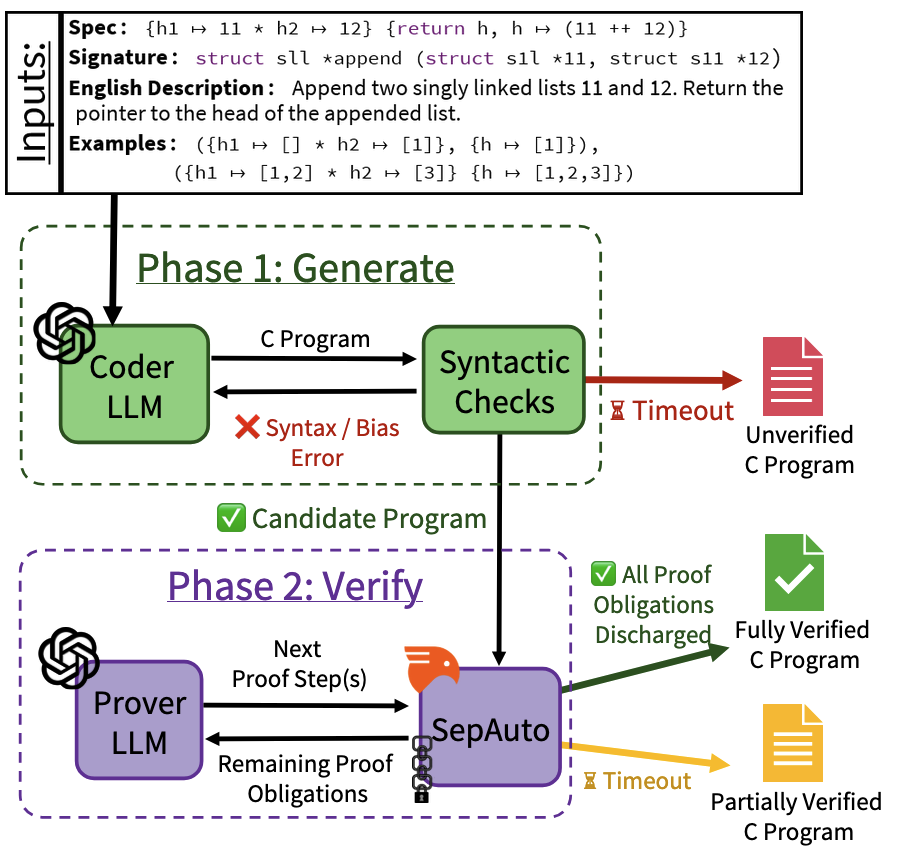}
  \vspace{-0.25em}
  \caption{Overview of our approach}
  \label{fig:overview}
  \vspace{-1em}
\end{figure}

\autoref{fig:overview} depicts the high-level workflow of our synthesis pipeline,
which is divided into two phases: generation and
verification. Our system takes four inputs that describe the target
function: its signature, its separation logic specification, a natural
language description of its behavior, and a couple of input-output
examples. In the first phase, these inputs are used to prompt a
\textit{coder LLM} to synthesize a candidate program. The candidate
program is then checked for syntax errors and conformance with our
syntactic biases.  If either check fails, the coder LLM is re-prompted
upto a threshold.

Otherwise, our pipeline proceeds to the second phase, which attempts
to verify that the candidate program meets its separation logic
specification using VST. This phase works by iteratively attempting to
discharge a set of outstanding proof obligations; this set initially
consists of only the top-level statement of correctness for the
candidate. Each iteration of this phase first applies \sepauto{}, our
custom automation proof tactic, to the current set of proof
obligations, producing a new (possibly empty) set of obligations. If
all obligations are satisfied, the system returns the verified program
and the complete proof script. Otherwise, a \textit{prover LLM} is
asked to either a) suggest tactics that will make progress on the
remaining obligations or b) identify a goal as unsolvable. These
suggestions are used to further refine the current proof script, and
the loop continues. To ensure termination, this phase places an upper
bound on the number of iterations; if this threshold is reached, the
candidate program and the current (incomplete) proof script are
returned.

The remainder of this section describes the generation and
verification phases of our pipeline in more detail, using the input
from \autoref{fig:overview} as a running example.

\subsection{Phase 1: Program Generation}

\begin{figure}[!htb]
\begin{center}
\begin{mdframed}[linecolor=black, linewidth=0.5pt,innerleftmargin=4pt, innerrightmargin=4pt, innertopmargin=1pt,innerbottommargin=1pt]
\begin{lstlisting}[basicstyle=\scriptsize\sffamily, escapeinside=||]
|Translate the given specification to a C program. Only include the C program in the content. No need to include a main function in the translated C program. There should not be loops in the program. All loops must be replaced by recursion. The only helper functions permitted are the ones provided under 'Helper functions:'. There should not be any novel helper functions used in the program. Generate recursive code if and only if non-loopy code generation is not possible. The code must compile using \textit{CLightGen}. The user provides the specification,
followed by the function name, signature, English description, and two input output examples of the function behaviour.|

Here are a couple of examples of how generated C
programs look given the input specifications:
[ |\textcolor{blue}{swap-spec}|, void swap(int *a, int *b) {..},..] (|\textcolor{gray}{elided for spaces}|)

Please provide the specifications as asked below:
Specification: |\textbf{$\langle$Spec$\rangle$}|
Function Name: |\textbf{$\langle$Name$\rangle$}|
Function Signature: |\textbf{$\langle$Signature$\rangle$}|
English Description: |\textbf{$\langle$English$\rangle$}|
Input Output Examples: |\textbf{$\langle$Example1$\rangle\langle$Example2$\rangle$}|
Helper Functions: |\textbf{$\langle$Signature, English, Spec$\rangle$}|
\end{lstlisting}
\end{mdframed}
\end{center}
\vspace{-0.5em}
\caption{The initial prompt template for the coder LLM}
\label{fig:initialGenPrompt}
\end{figure}


\synver{} generates an initial candidate C program by querying the
coder LLM using the prompt template shown in
\autoref{fig:initialGenPrompt}. This prompt includes a few examples of
input specifications along with their expected output C programs. The
placeholders in the prompt, e.g., \textbf{$\langle$Spec$\rangle$} and
\textbf{$\langle$Name$\rangle$} are then instantiated with the
arguments provided to \synver{}. The top of \autoref{fig:overview}
includes example inputs for synthesizing a C function that appends two
singly linked lists. \synver{} first checks if the program is a valid
C program by attempting to compile it. If compilation fails, \synver{}
re-prompts the coder LLM using the first template in
\autoref{fig:failedGenPrompt}, instantiating its placeholder with the
compilation error.

\begin{figure}[!htb]
\begin{center}
\begin{mdframed}[linecolor=black, linewidth=0.5pt, skipabove=2pt,
    skipbelow=2pt, innerleftmargin=2pt, innerrightmargin=2pt, innertopmargin=2pt,innerbottommargin=2pt]
\begin{lstlisting}[basicstyle=\scriptsize\sffamily, escapeinside=||]
|The program generated has the syntax error: $\langle$\textbf{Error}$\rangle$. Please re-generate the program such that it compiles with \textit{CLightGen}.|
\end{lstlisting}
\end{mdframed}
\vspace{-1em}
\begin{lstlisting}[basicstyle=\scriptsize\sffamily, escapeinside=||]
----------------------------------------
\end{lstlisting}
\vspace{-0.7em}
\begin{mdframed}[linecolor=black, linewidth=0.5pt, skipabove=2pt,
    skipbelow=2pt,innerleftmargin=2pt, innerrightmargin=2pt, innertopmargin=2pt, innerbottommargin=2pt]
\begin{lstlisting}[basicstyle=\scriptsize\sffamily, escapeinside=||]
|The program generated violated the syntactic bias: $\langle$\textbf{Bias-Type}$\rangle$. Please re-generate the program adhering to the syntactic biases.|
\end{lstlisting}
\end{mdframed}
\end{center}
\vspace{-0.5em}
\caption{Templates used to re-prompt the coder LLM}
\label{fig:failedGenPrompt}
\vspace{-0.5em}
\end{figure}


As \autoref{sec:overview} discussed, in VST (and Rocq more broadly)
proofs of program correctness are typically constructed
\emph{interactively}. VST provides a set
of specialized tactics for working with the proof obligations that
arise when reasoning about C programs in its logic, some of which,
e.g. \lstinline{forward_if}, and \lstinline{forward_loop}, can require
additional input from the user driving the process. To limit the need
for user interaction when reasoning about the programs it synthesizes,
\synver{} places two key restrictions, or \emph{biases}, on the syntax
of candidate programs.

\paragraph{No calls to functions without specifications} VST requires
logical specifications for any functions used by the program being
verified. Function specifications are stored in a context; if a
function is missing from the context, a user must manually supply its
specification. To avoid this situation, the initial generator prompt
(\autoref{fig:initialGenPrompt}) instructs the LLM to not to introduce
any intermediate helper functions, and \synver{} checks that a
candidate program only calls functions whose specifications are
included in the global context. As an example, \lstinline{swap} in
\autoref{fig:examplerecIter}, is not a candidate for verification, as
it introduces and calls the helper function \lstinline{add}, which
lacks a corresponding specification.  When this occurs, \synver{}
re-prompts the coder LLM using the second template in
\autoref{fig:failedGenPrompt}, filling in the
\textbf{$\langle$Bias-Type$\rangle$} placeholder instantiated with a
message noting that \lstinline{add} lacks a specification.

\paragraph{No loops} As is standard in program logics, reasoning about
loops in VST requires users to supply a \emph{loop invariant}. When
dealing with separation logic assertions, such invariants can be quite
involved, often requiring complex operators like the separating
implication~\cite{xiao2023hyperwand}. Recursive function calls, in
contrast, can reuse the function's top-level specification and do not
require additional user input. Thus, the prompt used by \synver{}
stipulates that the generated program should avoid loops and use
recursion instead. Including this restriction in the prompt causes the
coder LLM to generate the program on the left of
\autoref{fig:examplerecIter}: while semantically equivalent to the
function in \autoref{fig:app+Impl}, reasoning about this version does
not require any logical specifications on top of the one in
\autoref{fig:app+Impl}.  When the coder LLM generates a program with
loops, \synver{} will re-prompt it using the second template in
\autoref{fig:failedGenPrompt}, filling in the
\textbf{$\langle$Bias-Type$\rangle$} placeholder with a message to
instructing the LLM to avoid using loops.

\begin{figure}[!htb]
  \centering
  \begin{tabular}{p{.5\linewidth}|p{.35\linewidth}}
    \begin{cprogl}
sll *append (sll *h1,
             sll *h2)
{
 if (h1 == NULL)  {
  return h2;
 } else {
  h1->next =
   append (h1->next, h2);
 }
 return h1;
}
\end{cprogl}
    &
\begin{cprogl}
void add (int *x){
  *x= *x + 1;
}
void swap (int *x,
           int *y){
 int a = *x;
 int b = *y;
 if(a < b){
  *x= b;
  *y = a;
 } else {
  *y = a;
 }
 add (x);
}
\end{cprogl}
  \end{tabular}
  \caption{A recursive (i.e., bias-correct) version of
    \cinline{append}, and a bias-incorrect \cinline{swap} program which
    calls a helper function without a specification}
  \label{fig:examplerecIter}
  \vspace{-1em}
\end{figure}

\subsection{Phase 2: Program Verification}


After it has generated a candidate program that meets these two
restrictions, \synver{} attempts to verify that program is correct via
\GenProof{} (\autoref{alg:repAlg}). This algorithm takes as input a
candidate program $p$, target pre- and postconditions $P$ and $Q$, and
a bound on the number of interactions with the prover LLM
$limit$. \GenProof{} either returns a complete proof script showing
that $\vdash \{P\} ~p~\{Q\}$ in VST, or as much of the proof it was
able to complete within the interaction bound. At a high level,
\GenProof{} mimics the standard proof development process in which a
developer examines the current goal to decide the next proof step,
tells the theorem prover to process the corresponding tactic, and then
repeats this loop until no subgoals remain. Alongside the current
proof script, \GenProof{} maintains a set of unsolved proof
obligations as a tree, $ptree$, where each node represents a goal and
its children correspond to subgoals generated by a tactic
application. Intuitively, the leaves of $ptree$ represent the goals
remaining after processing the current proof script; the proof is
completed when $ptree$ is empty.

\begin{algorithm}[!t]
  \scriptsize \DontPrintSemicolon
  \Procedure{\GenProof($C_p, P, Q, limit$)}{
    \Params{$C_p$: candidate C program\\
      $P, Q$: target pre- and postconditions} %
    \Output{\texttt{Complete Proof} or \texttt{Partial Proof}} %
    $\mathit{ptree} \leftarrow $
    \initProofTree{\sepauto{}(\{$~\vdash\{P\}\;C_p\;\{Q\}$\})} \; %
    $\mathit{curGoal} \leftarrow$ \nextGoal{$ptree$} \;%
  $\mathit{curPrompt} \leftarrow$ \InitialProverPrompt{$C_p$, $P$,
    $Q$, $curGoal$} \;
  \For{$1 \ldots limit$}{ %
    \lIf{$\mathit{curGoal} = \bot$}{\Return{$ptree$} \Comment*[r]{Complete proof}}%
    \Switch{\GenTacticsOrUnsolvable{$\mathit{curPrompt}$}}{
      \Case{\textsc{Unsolvable}}{
        $\mathit{curPrompt} \leftarrow$ \UnsolvablePrompt{$curGoal$}\;  %
        $curGoal \leftarrow$ \Parent{$curGoal$} \; %
        \DeleteAllChildren{$curGoal$} \; %
      }
      \Case{\textsc{Try} $C_\mathit{tac}$}{
          $\mathit{resp} \leftarrow \emptyset$ \;
          \While{$C_\mathit{tac} \neq \emptyset$} {
            $\mathit{curTactic} \leftarrow$ \Pop{$C_\mathit{tac}$}\;%
            \Switch{\takeSteps{($g$, $\mathit{curTactic}$)}}{
              \Case{\textsc{Progress} $\mathit{subGoals}$}{
                $\mathit{subGoals'} \leftarrow$ \sepauto{}($\mathit{subGoals}$)\;
                \If{$\mathit{subGoals'} \neq \emptyset$}{
                  $\mathit{curGoal} \leftarrow$\AddGoals{$\mathit{ptree}$, $\mathit{subGoals'}$}\;
                }
                \Else{
                  $\mathit{curGoal} \leftarrow$ \RemoveGoal{$\mathit{ptree}$, $\mathit{curGoal}$}\;
                }
                $\mathit{curPrompt} \leftarrow$ \SuccessPrompt{$\mathit{curGoal}$}\;
                \textbf{break}\;
              }
              \Case{\textsc{Fail} $\mathit{msg}$}{
                $\mathit{resp} \leftarrow \mathit{resp} \cup \{(curTactic, \mathit{msg})\}$\;
                 \lIf{$C_\mathit{tac} = \emptyset$}{
                 \; $\mathit{curPrompt} \leftarrow$
                 \FailPrompt{$\mathit{curGoal}, \mathit{resp}$}}
              }
            }
          }

        }
      }



   }
   \Return{$ptree$} \Comment*[r]{Partial proof}
 }
\caption{Proof Generation}
\label{alg:repAlg}
\end{algorithm}

\GenProof{} first uses \sepauto{} (\autoref{alg:verifAlg}), a custom
tactic that simplifies and discharges VST-specific proof obligations,
to simplify the top-level goal. It then uses any subgoals generated by
\sepauto{} to construct an initial proof tree (line 2) and chooses one
of these goals to work on next (line 3). \GenProof{} uses the selected
subgoal to construct the initial prompt for the prover LLM (line
4). This prompt uses the template shown in
\autoref{fig:initialProverPrompt} to suggest a prioritized list of
five tactics that could help resolve this goal. This template begins
with a set of helper definitions and lemmas and some example proofs in
VST; its placeholders are instantiated with the candidate program and
its CompCert AST (\textbf{$\langle$Cand-Prog$\rangle$} and
\textbf{$\langle$C-AST$\rangle$}, its formal specification and
top-level statement of correctness
(\textbf{$\langle$VST-Spec$\rangle$} and
\textbf{$\langle$Thm-Stmnt$\rangle$}), and the current proof
obligation (\textbf{$\langle$Cur-Goal$\rangle$}).

\begin{figure}[!htb]
\vspace{-1em}
\begin{center}
\begin{mdframed}[linecolor=black, linewidth=0.5pt,
     innerleftmargin=4pt, innerrightmargin=4pt, innertopmargin=2pt,innerbottommargin=2pt]
\begin{lstlisting}[basicstyle=\scriptsize\sffamily, escapeinside=||]
|You are an expert in Coq, specifically in Separation Logic and the Verified Software Toolchain module. Please help me prove the correctness of CompCert C programs in VST (version 2.14) and Coq (version 8.19.2), against the specification in VST. It is recommended to use VST Floyd tactics like forward, forward\_if, entailer!! etc. to advance most of the proofs in this setting. Here are some additional definitions and lemmas you may need to use that are not included in the VST codebase:|
|[\textcolor{blue}{Definition t\_list := ..,  Lemma nullBST := BST E ,..}] (\textcolor{gray}{elided for spaces})|

|Here are 4 examples of how VST proofs look like given the CompCert AST and specification:|
|[\textcolor{blue}{Definition swap\_spec : ident * funspec := .., Definition f\_swap := { .. } ,..}] (\textcolor{gray}{elided for spaces})|

|Given the C-code: $\langle$\textbf{Cand-Prog}$\rangle$, corresponding CompCert AST:  $\langle$\textbf{C-AST}$\rangle$ and VST specification:  $\langle$\textbf{VST-Spec}$\rangle$, your task is to prove the lemma: $\langle$\textbf{Thm-Stmnt}$\rangle$ by specifying a set of up to 5 tactics to advance the current goal, in order of highest probability of success. The interpreted state would then be returned back to you, and you will predict the next set of tactics, till a fixpoint is reached, or the proof is completed. All tactics you predict, must be only relevant to the current goal.|

|Current Goal: $\langle$\textbf{Cur-Goal}$\rangle$|
\end{lstlisting}
\end{mdframed}
\end{center}
\vspace{-0.5em}
\caption{Template used to construct initial prompt to the prover LLM}
\vspace{-0.25em}
\label{fig:initialProverPrompt}
\end{figure}

\GenProof{} next enters a loop that attempts to iteratively verify the
candidate program meets its specification. This loop continues until
all proof obligations are discharged or the interaction limit with the
prover LLM has been reached. In the latter case, \GenProof{} returns
the partial proof up to that point (line 30).  If no goals remain,
\GenProof{} returns the complete proof (line 6). Each iteration of the
loop first queries the prover LLM with the current prompt. The prover
LLM may flag the current goal as unsolvable, which can happen when an
earlier iteration chose the wrong tactic, e.g., applying a lemma whose
assumptions are not provable from the current set of hypotheses. When
this occurs, \GenProof{} backtracks to the parent of the current goal
in $\mathit{ptree}$, i.e., the goal that spawned the current
(unsolvable) proof obligation (lines 10-12), and attempts a different
proof strategy.

\begin{figure}[!htb]
\begin{center}
\begin{mdframed}[linecolor=black, linewidth=0.5pt,innerleftmargin=4pt, innerrightmargin=4pt, innertopmargin=2pt,innerbottommargin=2pt]
\begin{lstlisting}[basicstyle=\scriptsize\sffamily, escapeinside=||]
|Given the goal: $\langle$\textbf{current-goal}$\rangle$, predict the next set of tactics to advance the goal.
You must predict at most five tactics, all of whom only advance this goal by one step, in order of highest probability of success. If the goal cannot be solved, please respond with \textit{Unsolvable}.|
\end{lstlisting}
\end{mdframed}
\end{center}
\vspace{-0.5em}
\caption{Prompt template used after a successful tactic application, or after backtracking one level up the proof tree}
\label{fig:successProverreprompt}
\vspace{-0.5em}
\end{figure}

Alternatively, \GenProof{} calls the \takeStep{} subroutine to each
tactic proposed by \GenTacticsOrUnsolvable (line 17). This subroutine
decides whether to apply the current tactic $\mathit{curTactic}$ to
$\mathit{curGoal}$. First, \takeStep{} checks if the current tactic
has been tried before. If not, it then asks Rocq to apply
$\mathit{curTactic}$ to $\mathit{curGoal}$; if Rocq reports an error
or the goal is unchanged, the tactic is rejected. Otherwise,
\takeStep{} examines the subgoals that result from applying
$\mathit{curTactic}$. If any of the new subgoals are equivalent to an
unsolved goal in $\mathit{ptree}$, there is a cycle in the current
proof, and \takeStep{} rejects $\mathit{curTactic}$. Next, if the
conclusion of any of the new subgoals is more than twice the size of
the conclusion of $\mathit{curGoal}$, $\mathit{curTactic}$ is
rejected. When neither of these situations occur, \GenProof{} accepts
the current tactic, and adds the resulting subgoals to
$\mathit{ptree}$ (line 21). If no subgoals have been generated, the
current proof obligation has been discharged and it is removed from
$\mathit{ptree}$, as are any of its ancestors that have been solved
(line 23). In both cases, the prompt for the prover LLM is updated
using the template in \autoref{fig:successProverreprompt} instantiated
with a new subgoal (line 24), and the main loop continues. If all
proposed tactics are rejected, the prover LLM is prompted again using
the fallback template in \autoref{fig:failureProverreprompt} (line
29), which includes information about the failing tactics.


\begin{figure}[!htb]
\begin{center}
\begin{mdframed}[linecolor=black, linewidth=0.5pt,innerleftmargin=4pt, innerrightmargin=4pt, innertopmargin=2pt,innerbottommargin=2pt]
\begin{lstlisting}[basicstyle=\scriptsize\sffamily, escapeinside=||]
|The tactics: $\langle$\textbf{tactic1,..,tactic5}$\rangle$ failed because of the following reasons: $\langle$\textbf{reason1,...,reason5}$\rangle$. Please re-generate the tactics for the current goal: $\langle$\textbf{current-goal}$\rangle$. If the goal cannot be solved, please respond with \textit{Unsolvable}.|

|The proof generated by you so far is: $\langle$\textbf{all-tactics-tried}$\rangle$. The correct proof generated so far is: $\langle$\textbf{current-proof}$\rangle$.|
\end{lstlisting}
\end{mdframed}
\end{center}
\vspace{-0.5em}
\caption{Prompt template used after application of all predicted tactics resulted in failure}
\vspace{-0.25em}
\label{fig:failureProverreprompt}
\end{figure}

\autoref{alg:verifAlg} presents \sepauto{}, the symbolic reasoning
subroutine that \GenProof{} uses to simplify and solve VST-specific
proof obligations. This function maintains sets of outstanding and
unresolved goals, $G_o$ and $G_u$ respectively. \sepauto{} attempts to
iteratively simplify or solve each goal in $G_o$ using a combination
of custom and VST-provided tactics. Goals that cannot be simplified or
solved are added to $G_u$, which \sepauto{} returns when no goals
remain in $G_o$. In each iteration its main loop, \sepauto{}
identifies the current goal as either a Hoare triple (lines 7-20) or a
side condition (lines 21-25). In the former case, \sepauto{} uses the
shape of the program in the triple to decide how to proceed.

\begin{algorithm}[!t]
  \scriptsize
  \DontPrintSemicolon
  \Procedure{\sepauto{}($G_o$)}{
  \Params{
    $G_o$: Initial set of outstanding goals}
  \Output{Set of unresolved goals}
  $G_u \leftarrow \emptyset$ \Comment*[r]{Unresolved goals}
   \While{$G_o \neq \emptyset$} {
    $g \leftarrow $ \firstGoal{$G_o$} \;
    $G_o \leftarrow G_o \setminus \{g\}$ \;
    \Switch{$g$}{
      \Case{H $\vdash$ \{P\} \lstinline[basicstyle=\ttfamily\scriptsize]{if b then c1 else c2} \{Q\}}{
        $G_o \leftarrow G_o \cup \{\texttt{forward\_if(g)}\}$
      }
      \Case{H $\vdash$ \{P\} \lstinline[basicstyle=\ttfamily\scriptsize]{(if b then c1 else c2);c3} \{Q\}}{
        $g \leftarrow $ H $\vdash$ \{P\}
        \lstinline[basicstyle=\ttfamily\scriptsize]{if b then c1;c3 else c2;c3} \{Q\}\;
        $G_o \leftarrow G_o \cup \{\texttt{forward\_if}(g)\}$
      }
      \Case{H $\vdash$ \{P\} \lstinline[basicstyle=\ttfamily\scriptsize]{f(x1, x2, ..., xn)} \{Q\}}{
        $f_p \leftarrow $ \inferCallParams($g$) \;
        $g' \leftarrow $ \{\texttt{forward\_call}($g, f_p$)\} \;
        \lIf*{$g' \neq g$}{
          $G_o \leftarrow G_o \cup \{g'\}$\Comment*[r]{Progress}
        }
        \lElse{
          $G_u \leftarrow G_u \cup \{g'\}$
        }
      }
      \Case{H $\vdash$ \{P\} \lstinline[basicstyle=\ttfamily\scriptsize]{c} \{Q\}}{
        $g' \leftarrow $ \texttt{forward\_withauto($g$)} \;
        \lIf*{$g' \neq g$}{
          $G_o \leftarrow G_o \cup \{g'\}$\Comment*[r]{Progress}
        }
        \lElse{
          $G_u \leftarrow G_u \cup \{g'\}$
        }
      }
      \Case{H $\vdash$ \_ \Comment*[r]{Side Condition}} {
        $g' \leftarrow $ \texttt{resolve\_withauto($g$)} \;
        \lIf*{$g' \neq g$}{
          $G_o \leftarrow G_o \cup \{g'\}$\Comment*[r]{Progress}
        }
        \lElse{
          $G_u \leftarrow G_u \cup \{g'\}$
        }
      }
    }
  }
\Return{$G_u$}
}
\caption{\sepauto{} - Proof automation for VST}
\label{alg:verifAlg}
\end{algorithm}

If the goal is a hoare triple involving a conditional statement (lines
7-8), \sepauto{} applies VST's built-in \tactic{forward\_if} tactic,
creating new subgoals for the \cinline{then} and \cinline{else}
branches.  If the conditional is sequenced with another statement,
i.e., \cinline{(if b then c1 else c2); c3}, however, applying
\tactic{forward\_if} would create a third goal for the trailing
statement \cinline{c3}. In this case, the tactic expects the user to
provide a precondition capturing the program state after the
conditional is executed. Since \sepauto{} is meant to be fully
automatic, it first rewrites a goal of this form into an equivalent
one by moving \cinline{c3} inside each of the branches. This allows
\tactic{forward\_if} to be applied as before, without any additional
input (lines 10-11).


The built-in VST tactic for function calls, \tactic{forward\_call},
expects users to explicitly provide arguments for each parameter of
the called function. To automate this step, \sepauto{} uses a custom
subroutine \inferCallParams that extract these arguments from the
current goal (lines 13-14). Alternatively, when the current goal is a
Hoare triple that is not covered by one of these three cases, e.g. it
is \lstinline{c1; c2} or %
\lstinline{x := a}, \sepauto{} applies \tactic{forward\_withauto}, an
enhanced version of VST's \tactic{forward} tactic. This custom tactic
tries to automatically discharge side conditions related to memory
safety, e.g., ensuring that a pointer is not null before it is
dereferenced. To discharge these sorts of side conditions,
\tactic{forward\_withauto} combines the current set of assumptions with
a custom library of helper lemmas. As a simple example, when reasoning
about line 11 of \autoref{fig:app+Impl}, \sepauto{} needs to ensure
that \lstinline|current| is non-null, which it does by rewriting the
goal using the lemma \tactic{listrep\_nonnull}. Applying this lemma
generates a new subgoal, similar to the proof on line 9 of
\autoref{fig:loopInvariant}, which \sepauto{} attempts to solve
automatically. If \tactic{forward\_withauto} cannot make progress, the
current goal is added to $G_u$. Finally, \sepauto{} uses a custom
\tactic{resolve\_withauto} tactic to resolve goals that are side
conditions (line 23) using a combination of standard, e.g.,
\tactic{list\_solve} and \tactic{rep\_lia}, and VST-specific tactics,
e.g., \tactic{entailer}. Any goals that are not completely solved by
\tactic{resolve\_withauto} are added to $G_u$ (line 25).



In summary, \synver{} implements a two-phase approach to synthesizing
formally verified C programs from high-level specifications. The first
phase uses a coder LLM to generate a program whose shape facilitates
automated verification. The second phase then attempts to build a
formal proof that the program satisfies a user-provided separation
logic specification using VST. To automate this proof, \synver{} uses
a complementary combination of symbolic reasoning (\sepauto{}) and
LLM-aided proof generation (\GenTacticsOrUnsolvable). The former
handles proof obligations generated by VST, while the latter handles
goals that \sepauto{} cannot.

\section{Evaluation}
\label{sec:exp}
Our evaluation investigates three key research questions about our
approach:
\begin{itemize}
\item {\bf RQ1}: How effective is \synver{}? Is it able to
  automatically generate fully verified C programs for a diverse set
  of synthesis tasks?
\item{\bf RQ2}: How much does each component of our prompt contribute
  to its ability to generate bias-correct programs that satisfy their
  separation logic specification?
\item{\bf RQ3}: How does \GenProof{} compare to
  other proof automation approaches?
\end{itemize}

All of our experiments were carried out on an Apple M2 Max Macbook Pro
with 32GB RAM, except for our Rango evaluation \cite{rangoICSE2025},
which was carried out on a NVIDIA 5500 GPU with 24GB RAM. \synver{}
uses \coderLLM{} for both its coder and prover LLMs, and limits the
number of LLM interactions in the first and second phases to 10 and
50, respectively.  

\subsection{Benchmark Construction}
To evaluate our approach, we developed a suite of specifications for a
set of programs of varying complexity. Each of the benchmarks used in
our evaluation falls into three distinct categories. The first
category (\textbf{Basic}) consists of programs that only use simple
built-in datatypes, e.g., \cinline{int} and \cinline{char}, and
arrays. This category includes 19 programs of varying complexity that
were adapted from a collection of formally verified Dafny
programs~\cite{LLMDafny2024}. The second set of benchmarks
(\textbf{Heap}) includes 24 programs that manipulate heap-allocated
data structures like singly linked lists and trees; these were drawn
from the evaluation suite of a prior deductive
synthesizer~\cite{suslikICFP21}. The final class (\textbf{API})
consists of adaptations of standard textbook algorithms~\cite{clrs},
and is made up of 5 programs that make function calls and use
structured datatypes, e.g., arrays, lists, and trees. To ensure that
programs in this class can be automatically verified, each callable
function is equipped with a formal specification. The specifications
for the API manipulating programs were derived from their textbook
specification by a verification expert.




\begin{table}[t!]
  \centering
  \caption{The results of \synver{} on the benchmarks it was able to
    completely verify. The three groups of rows correspond to the
    \textbf{Basic}, \textbf{Heap}, and \textbf{API} categories,
    respectively. The \textbf{Rec} column indicates whether the target
    program makes a recursive call, and \textbf{LoC} gives the length
    of the generated program. \textbf{GP} reports the number of proof
    obligations generated by the initial call to \sepauto{} --- a
    value of 0 means \sepauto{} was able to fully verify the candidate
    program.  \textbf{LoP} is the number of lines in the generated
    proof script.  \textbf{PT} and \textbf{GT} report the time needed
    to fully verify and generate a program, respectively. \textbf{MS} gives the number
    of tactics discarded and the number of times \GenProof{} backtracked.
    \textbf{TE} gives the total number of tactics \takeStep{}
    evaluated. }
  \label{fig:evaluation-succ}
  \footnotesize
  \setlength{\tabcolsep}{4pt}    
  \renewcommand{\arraystretch}{1.0}
  \begin{adjustbox} {max width=\columnwidth,keepaspectratio}
  \begin{tabular}{|l|c|c|c|c|c|c|c|c|r}
    \hline
    \textbf{Benchmark}                    & \textbf{Rec}  & \textbf{LoC}     &  \textbf{GP}      & \textbf{LoP}  & \textbf{PT} & \textbf{GT} & \textbf{MS}  & \textbf{TE} \\
    \hline
    isEven                                & \NoOK         & 7                  & 2                              & 5          & 2m 54s             & 9.29s       & 0          & 4 \\
    getElement                            & \NoOK         & 11                 & 3                              & 12         & 10m                & 9.08s       & 15         & 26 \\
    isDivBy11                             & \NoOK         & 7                  & 2                              & 3          & 1m 36s             & 7.05s       & 2          & 4  \\
    minTwo                                & \NoOK         & 7                  & 0                              & 1          & 1.8s               & 3.67s       & 0          & 0  \\
    mulTwo                                & \NoOK         & 3                  & 0                              & 1          & 1.2s               & 4.24s       & 0          & 0 \\
    minThree                              & \NoOK         & 9                  & 0                              & 1          & 4.4s               & 8.20s       & 0          & 0  \\
    lastDigit                             & \NoOK         & 3                  & 0                              & 1          & 1.5s               & 5.65s       & 0          & 0 \\
    nIsGreater                            & \OK           & 12                 & 3                              & 26         & 53m               & 17.02s      & 53          & 78  \\
    arrayModify                           & \OK           & 9                  & 0                             & 1          & 3.6s                & 7.95s       & 0          & 0  \\
    addBy1                                & \OK           & 9                  & 1                             & 2          & 29.40s              & 8.70s       & 0          & 4  \\
    allSame                               & \OK           & 11                 & 3                             & 21         & 14m 48s             & 15.86s      & 7          & 27  \\
    \hline \hline
    swap                                  & \NoOK         & 5                  & 0                              & 1         & 2.6s             & 5.26s      & 0            & 0  \\
    swapdAdd                              & \NoOK         & 7                  & 0                              & 1         & 2.3s             & 5.80s      & 0            & 0  \\
    swapIf                                & \NoOK         & 11                 & 2                              & 7         & 4m               & 13.70s     & 4            & 10  \\
    assignX                               & \NoOK         & 3                  & 0                              & 1         & 2s               & 4.85s      & 0            & 0  \\
    assignYAdd                            & \NoOK         & 3                  & 0                              & 1         & 3.5s             & 6.66s      & 0            & 0  \\
    listLength                            & \OK           & 8                  & 2                              & 6         & 2m               & 8.92s      & 0            & 5  \\
    listFree                              & \OK           & 9                  & 1                              & 3         & 43.1s            & 8.61s      & 0            & 2 \\
    isListEmpty                           & \NoOK         & 11                 & 2                              & 10        & 4m 30s           & 7.77s      & 2            & 11  \\
    listAppend                            & \OK           & 11                 & 2                              & 7         & 2m 54s           & 7.70s      & 0            & 6  \\
    listInsBeg                            & \NoOK         & 6                  & 1                              & 4         & 59s              & 10.36s     & 0            & 3  \\
    listDelEnd                            & \OK           & 12                 & 4                              & 15        & 32m 30s          & 11.45s     & 13           & 27  \\
    listLookup                            & \OK           & 9                  & 3                              & 24        & 10m 54s          & 11.20s     & 2            & 25  \\
    listCopy                              & \OK           & 10                 & 2                              & 9         & 3m 42s           & 10.19s     & 0            & 8  \\
    listInsEnd                            & \OK           & 11                 & 2                              & 9         & 3m 48s           & 12.00s     & 0            & 8  \\
    listFilter                            & \OK           & 15                 & 3                              & 14        & 6m 24s           & 14.55s     & 0            & 13  \\
    listAdd1                              & \OK           & 9                  & 2                              & 5         & 1m 36s           & 7.72s      & 0            & 4  \\
    listDelBeg                            & \NoOK         & 13                 & 2                              & 20        & 29m 54s          & 12.65s     & 41           & 60  \\
    bstFree                               & \OK           & 8                  & 0                              & 1         & 3.6s             & 6.70s      & 0            & 0  \\
    bstLookup                             & \OK           & 15                 & 3                              & 35        & 32m 12s          & 14.19s     & 20           & 54  \\
    bstInsert                             & \OK           & 24                 & 4                              & 22        & 17m 24s          & 16.81s     & 7            & 28  \\
    bstMinValue                           & \OK           & 7                  & 3                              & 11        & 6m 36s           & 7.90s      & 0            & 10  \\
    \hline \hline
    bstSkewed                             & \NoOK        & 3                   & 2                             & 12        & 7m               & 22.77s     & 6            & 17  \\
    popHighest                            & \NoOK        & 5                   & 3                             & 8         & 3m 24s           & 18.87s     & 2            & 9  \\
    \hline
  \end{tabular}
  \end{adjustbox}
\end{table}

\begin{table}[t!]
  \centering
  \caption{The results for the benchmarks \synver{} was only able to
    partially verify. \textbf{SP} reports how many of the initial
    subgoals from \sepauto{} that \GenProof{} completely solved.}
  \label{fig:evaluation-fail}
  \footnotesize
  \setlength{\tabcolsep}{4pt}    
  \renewcommand{\arraystretch}{1.0}
  \begin{adjustbox} {max width=\columnwidth,keepaspectratio}
  \begin{tabular}{|l|c|c|c|c|c|c|c|c|c|r}
    \hline
    \textbf{Benchmark}                    & \textbf{Rec}  & \textbf{LoC}     & \textbf{GP}  &  \textbf{SP}      & \textbf{LoP}  & \textbf{PT} & \textbf{GT} & \textbf{MS}  & \textbf{TE} \\
    \hline
    checkSorted                           & \OK           & 19                 & 3                   & 2           & 34         & 1h 18m             & 19.23s      & 80          & 115 \\
    checkZ                                & \OK           & 11                 & 4                   & 1           & 17         & 1h                 & 9.54s       & 100         & 117  \\
    consecNums                            & \OK           & 11                 & 4                   & 3           & 26         & 1h 18m             & 18.02s      & 108         & 134  \\
    firstOddIndex                         & \OK           & 11                 & 4                   & 3           & 33         & 1h 24m             & 9.98s       & 76          & 109  \\
    arrayMember                           & \OK           & 9                  & 3                   & 1           & 12         & 1h 6m              & 7.75s       & 167         & 180  \\
    OddAtOdd                              &\OK            & 14                 & 5                   & 1           & 13         & 1h 6m              & 22.93s      & 134         & 152 \\
    lastPosition                          & \OK           & 11                 & 2                   & 1           & 5          & 40m 48s            & 14.87s      & 200         & 204  \\
    compArrays                            & \OK           & 11                 & 4                   & 3           & 41         & 2h 36m             & 15.69s      & 68          & 108  \\
    \hline \hline
    listArrayEq                           & \OK           & 16                 & 4                   & 1           & 31        & 1h 30m           & 13.78s     & 86            & 119  \\
    bstMinNode                            & \OK           & 10                 & 3                   & 0           & 34        & 4h 18m           & 10.35s     & 119           & 154  \\
    bstMinKey                             & \OK           & 7                  & 3                   & 2           & 42        & 1h 36m           & 6.15s      & 70            & 113 \\
    \hline \hline
    countValue                            & \NoOK        & 10                  & 5                   & 0           & 27        & 1h 42m           & 12.51s     & 138            & 164  \\
    bstDel                                & \OK          & 31                  & 6                   & 1           & 30        & 4h 24m           & 20.55s     & 127            & 156  \\
    addLast                               & \NoOK        & 12                  & 3                   & 0           & 28        & 2h               & 15.76s     & 106            & 133  \\
    \hline
  \end{tabular}
  \end{adjustbox}
\end{table}

\subsection{RQ1: Effectiveness of \synver{}}
\label{sec:RQ1}
\autoref{fig:evaluation-succ} and \ref{fig:evaluation-fail} presents
the results of \synver{} for each of the input specifications in our
benchmark suite. \synver{} was able to successfully generate programs
of varying lengths, ranging from 3 to 31 lines of code, with an
average length of 10 lines. On average, it took \coderLLM{} 11.23
seconds to produce a candidate program. In all cases, \coderLLM{} was
able to generate a syntactically valid candidate program meeting our
biases on the first try--- reprompting was never needed.

\autoref{fig:evaluation-succ} shows the results for the 70\% (34/48)
of our benchmarks that \GenProof{} was able to fully and automatically
verify. These include both simple programs (e.g., \texttt{mulTwo}
simply multiplies two integers) and more complex ones (e.g.,
\texttt{insertBST} is a recursive function that inserts an element
into a binary search tree and has a complex specification). Of these
34 programs, \GenProof{} was able to automatically verify 10 programs
with just its initial call to \sepauto{}. Since it did not have to
query the prover LLM, the time needed to verify each of these
benchmarks was quite short, under 5 seconds. For the remaining 24
programs, \GenProof{} produced proof scripts of varying lengths,
ranging from 2 to 35 proof lines, with an average length of 9, where
each line consists of a tactic suggested by \coderLLM{} followed by a
call to \sepauto{}. When generating 11 of these 24 proofs, \GenProof{}
did not discard any tactics or backtrack; for the remaining 13,
\GenProof{} did one of these actions an average 38.7\% of the time.
The total time needed to generate these proofs corresponds to the
number of tactics tried by \GenProof{}, and ranged from 1.2 seconds to
53 minutes, with an average time of roughly 7.5 minutes.

Based on a manual analysis, \synver{} generated correct programs for
each of the remaining benchmarks, even though \GenProof{} was only
able to partially verify these programs within its interaction
limit. These programs tend to be longer and have more complex
specifications than those \GenProof{} was able to fully
verify. \autoref{fig:evaluation-fail} shows the results for these 14
benchmarks. As the table shows, \takeStep{} discards many more tactics
and backtracks more often in these benchmarks, on average 80\% of the
time. This suggests that \GenProof{} spent much of its time on these
benchmarks exploring unproductive proof directions. As a consequence,
the proof generation time for these benchmarks was considerably slower
than the for the fully verified benchmarks, with an average time of
nearly 2 hours. Note that our current implementation of \GenProof{}
asks Rocq to reprocess the current proof script in its entirety each
time \takeStep{} tries a new tactic--- thus, the overall proving time
increases substantially with the number of tactics \takeStep{}
evaluates. This overhead could be substantially reduced by
implementing via a more incremental interaction loop with the theorem
prover.

A manual inspection of the proof scripts generated by \GenProof{}
suggests that \verifyLLM{} can effectively compensate for gaps in
\sepauto{}'s automation. As one example, \sepauto{} does not attempt
to instantiate existential variables, a key part of correctness proofs
in VST for functions with a non-\cinline{void} return
type. \verifyLLM{} was able to identify the right instantiation in all
but three of our benchmarks. This investigation also indicates that
\verifyLLM{} was able to effectively identify and apply lemmas that
help the proof make progress.


We also performed a manual analysis of the 14 proofs \GenProof{} could
only partially complete. Based on this analysis, we categorized each
program into one of four categories:
\begin{itemize}
\item \textbf{Faulty Suggestions}: The partial proofs for
  \cinline{checkZ}, \cinline{OddAtOdd} and \cinline{lastPosition}
  feature a large number of tactics (86\%, 88\% and 98\%) that were
  discarded due to Rocq-reported errors or failure to make progress.
\item \textbf{Cyclic Reasoning}: The partial proofs for
  \cinline{consecNums}, \cinline{checkSorted}, and
  \cinline{lastPosition} all repeatedly try sequences of tactics that
  modify the goal in some way before eventually arriving at the original
  goal, effectively not making progress. Incorporating better cycle
  detection in \takeStep{} could help ameliorate these sorts of
  failures.
\item \textbf{Superfluous Tactic Suggestions}: The partial proofs for
  \cinline{firstOddIndex}, \cinline{listArrayEq},
  \cinline{bstMinNode}, \cinline{countValue}, and \cinline{bstDel}
  repeated call unnecessary tactics that, e.g., perform superfluous case analysis or induction.
\item \textbf{Backtracking Failure}: The partial proofs for the three
  remaining benchmarks all incorrectly instantiate an existential
  variable at some point, but \GenProof{} is unable to either detect
  this, in the case of \cinline{addLast}, or it detects the problem
  much further later in the proof script and is unable to quickly
  revert to the point at which the flawed reasoning occured, in the
  case of \cinline{arrayMember} and \cinline{bstMinKey}. For these
  last two benchmarks, allowing \GenProof{} to revert to an arbitrary
  earlier point in the proof could help lead to better results.
\end{itemize}

\subsection{RQ2: Composition of the coder LLM Prompt}
As discussed in \autoref{sec:approach}, \GenProof{} is designed to be
applied to semantically correct programs that conform to a set of
syntactic biases. This section presents an ablation study of how much
the individual components of the prompt given to the coder LLM
contributes to the ability of \synver{} to generate programs meeting
those biases. For this experiment, we have constructed five variants
of the prompt from \autoref{fig:initialGenPrompt} and use these to
prompt \coderLLM{} for programs for the 43 specifications belonging to
the first two benchmark categories. The API benchmark is omitted from
this experiment because their prompt includes additional information
in the form of the additional functions (with specifications) that the
synthesized program is allowed to call.



\begin{table}[t!]
\centering
\caption{The result of using different variations of the coder prompt
  with \coderLLM{}. The first column (\textbf{Var}) lists the prompt
  variant, followed by five columns indicating which components of the
  prompt were included. The first four of these indicate whether the
  prompt includes explicit instructions to follow the syntactic biases
  (\textbf{Bias}), a separation logic specification (\textbf{Spec}), a
  natural language description (\textbf{Desc}), and input-output
  examples (\textbf{Ex}). The next column lists the kind of function
  name that was included in the prompt: Original uses the name from
  Tables \ref{fig:evaluation-succ} and \ref{fig:evaluation-fail};
  Verbose uses a long but meaningful function name; and Arbitrary is a
  random name with no relation to the function's intent.  The next two
  groups of columns list the number of programs meeting the syntactic
  bias (\textbf{B}) and the number of correct programs (\textbf{C})
  generated in response to each prompt variations for the
  (\textbf{Basic}) and (\textbf{Heap}) benchmark categories. }
  \label{fig:evaluation-prompt}
  \vspace{1em}
  \renewcommand{\arraystretch}{1.0}
   \begin{adjustbox} {max width=\columnwidth,keepaspectratio}
\begin{tabular}{|c|c|c|c|c|c|c|c|c|c|}
    \hline
    \multirow{2}{*}{\textbf{Var}} &
    \multirow{2}{*}{\textbf{Bias}} &
    \multirow{2}{*}{\textbf{Spec}} &
    \multirow{2}{*}{\textbf{Desc}} & 
    \multirow{2}{*}{\textbf{Ex}} & 
    \multirow{2}{*}{\textbf{Name}} & 
       \multicolumn{2}{c|}{\textbf{Basic} (19)}
      & \multicolumn{2}{c|}{\textbf{Heap} (24)} \\
    \cline{7-10} 
    & & & & & & \centering\textbf{B} & \centering\textbf{C}
      & \centering\textbf{B} & \centering\textbf{C} \tabularnewline
    \hline
    \textbf{P1} & \NoOK & \OK       & \OK      & \OK       & Original & 7     & 9     & 11     & 24 \\
    \textbf{P2} & \OK   & \NoOK     & \NoOK    & \NoOK & Verbose     & 19    & 18    & 24     & 22 \\
    \textbf{P3} & \OK   & \OK         & \NoOK    & \NoOK    & Verbose & 19    & 19    & 24     & 24 \\
  \textbf{P4} & \OK   & \NoOK     & \OK      & \NoOK     & Arbitrary    & 19    & 19    & 24     & 24 \\
  \textbf{P5} & \OK   & \OK         & \NoOK    & \NoOK     & Arbitrary& 19    & 19    & 24     & 20 \\
\hline
  \end{tabular}
  \end{adjustbox}
  \end{table}

  \autoref{fig:evaluation-prompt} reports the results of manually
  checking whether the resulting programs were correct and conformed
  to our biases.  Each of these variations investigates a different
  aspect of the coder prompt:
  \begin{itemize}
  \item \textbf{P1}: This variation investigates the shape of the
    programs that the coder LLM generates without additional
    instructions. Without explicit guidance, \coderLLM{} generates
    programs following our syntactic bias less than half of the time,
    using loops instead of recursion in all but 4 cases; of these, 3
    are tree programs that naturally admit recursive solutions.

  \item \textbf{P2 and P3}: When provided with only a descriptive
    function name and instructions about our biases, \coderLLM{}
    responds with a correct program for all but three of our
    specifications, suggesting that \coderLLM{} is particularly
    sensitive to this prompt component. The programs in the three
    failing cases--- \cinline{consecNumbers}, \cinline{assignX} and
    \cinline{assignYAdd}--- had ambiguous names, further supporting
    this hypotheses.  Providing additional semantic information in the
    form of separation logic specifications (\textbf{P3}), enables
    \coderLLM{} to generate all correct programs, suggesting that
    including specifications can help the LLM when only part of the
    target functionality is encoded in the function name.

  \item \textbf{P4}: This variant probes how well the LLM responds to
    informal specifications. In contrast to \textbf{P2} and
    \textbf{P3}, however, all of the natural language descriptions
    used in this experiment are able to completely capture the
    behavior of the target program.  \coderLLM{} is able to generate
    correct programs for all our benchmarks with this variant,
    suggesting that it can effectively interpret mathematically
    imprecise specifications.

  \item \textbf{P5}: This final prompt variant tests how well
    \coderLLM{} is able to interpret mathematically rigorous
    specifications written in separation logic. \coderLLM{} was less
    effective when provided with just these sorts of specifications,
    however, failing to generate correct programs for four of our
    benchmarks using this prompt. Three of these---
    \cinline{listAdd1}, \cinline{listAppend}, and
    \cinline{listDelEnd}--- make a copy of the input list, which is
    inconsistent with the target specification. The last incorrect
    program generated by \coderLLM{}, \cinline{bstFree}, fails to
    properly deallocate the target list, simply returning
    \cinline{null} instead.
  \end{itemize}

  Taken together, these results suggest that it is important to
  include explicit instructions about our syntactic biases. In
  addition, while \coderLLM{} can effectively interpret natural
  language components, it is less effective at interpreting more
  formal specifications of target program behaviors.

\subsection{RQ3: Alternative Proof Automation Strategies}
To evaluate the effectiveness of \synver{}'s approach to proof
automation, we have conducted a comparative study of \GenProof{} with
other proof automation approaches. Our set of alternative approaches
includes two state-of-the art learning-based theorem provers,
Tactician \cite{blaauwbroek2020tactician} and Rango
\cite{rangoICSE2025}, three simplified variants of \GenProof{}, and
SA++, an enhanced version of \sepauto{} equipped with additional
tactic-based proof automation. SA++ is meant to serve as a roofline
for how well purely symbolic proof automation can work for VST-style
proofs of program correctness. This tactic was developed by a
verification expert who first manual wrote proof scripts for a subset of
our benchmarks, and then generalized the high-level proof strategies
used into a \rocqinline{crush}-style tactic~\cite{CPDT}, a process
that took about 100 person-hours.

\begin{table}[!thb]
  \centering
  \caption{The results of using alternative proof automation
    approaches to verify a subset of candidate
    programs. \textbf{Framework} lists the prover, and the
    remaining columns report the number of benchmarks from the
    \textbf{Basic}, \textbf{Heap}, and \textbf{API} categories each
    framework was able to completely verify. SA only applies
    \sepauto{} to the top-level theorem. GP-SA-H is a limited version
    of \GenProof{} that only calls \sepauto{} on the top-level
    theorem, and tries to discharge all subsequent proof obligations
    using just the prover LLM, does not backtrack, and uses a version
    of \takeStep{} that does not filter tactics based on the size of
    the goals they generate. GP+SA-H is a variant of GP-SA-H that also
    applies \sepauto{} to new subgoals. SA++ is an enhanced version of
    \sepauto{} that is equipped with additional tactic-based
    automation. }
  \label{fig:evaluation-ablation-Prover}
  \renewcommand{\arraystretch}{1.0}
   \begin{adjustbox} {max width=\columnwidth,keepaspectratio}
   \begin{tabular}{|l|c|c|c|r}
    \hline
     \textbf{Framework}  & \textbf{Basic (18)} & \textbf{Heap (22)}  & \textbf{API (2)} \\
     \hline
     Rango                     & 0                & 3            & 0       \\
     Tactician                 & 2                & 4            & 0     \\
     \hline \hline
     SA                        & 5                & 5            & 0      \\
     GP-SA-H                   & 9                & 14           & 0     \\
     GP+SA-H                   & 9                & 18           & 0       \\
     \GenProof{}               & 11               & 19           & 0      \\
     \hline \hline
     SA++                      & 14               & 21           & 1       \\
     \hline
  \end{tabular}
  \end{adjustbox}
\end{table}

We divide the benchmarks used in these experiments into two
groups. The first group consists of 42 benchmarks and admits a total
ordering based on the number of proofs each approach was able to
completely solve (shown in \autoref{fig:evaluation-ablation-Prover}):
Rango $\subset$ Tactician $\subset$ SA $\subset$ GP-SA-H $\subset$
GP+SA-H $\subset$ \GenProof{} $\subset$ SA++.

In general, the two learning-based proof automation approaches
performed quite poorly, with Tactician finishing slightly more (6)
proofs than Rango (3). We attribute this to the specialized nature of
proofs of program correctness in VST: there are not many publicly
available examples of such proofs, and the training data for both
provers is thus unlikely to include such proofs; Tactician supports
on-the-fly learning, and is thus able to perform better.  We provide
the four examples of VST proofs used in our prompt to the prover LLM
when evaluating both tools, and Tactician seems to learn from these
examples.

Of the 42 benchmarks that admit a total order on approaches, the three
limited variants of \GenProof{} all perform better than Rango and
Tactician, but are only able to solve a subset of the proofs that
\GenProof{} and SA++ can. The proofs that only SA++ is able to solve
completely, all feature specifications with nested quantifiers and
boolean predicates; showing the corresponding programs correct
requires correctly instantiating these quantifiers. The other proof
generation approaches struggle with this task, generating a large
number of discarded tactics and repeatedly backtracking. We also note
there are 6 programs that none of the approaches are able to
completely verify; all of these have particularly complex
specifications, e.g., combination of nested quantifiers and boolean
operators, and the separating implication operator.

\begin{table}[!htb]
  \centering
  \caption{The results of verifying the 6 benchmarks not included in
    \autoref{fig:evaluation-ablation-Prover}. Results for Rango,
    Tactician, and SA are omitted, as all three are unable to
    completely verify any of these programs.  The columns after
    \textbf{Framework} correspond to these 6 benchmarks:
    \cinline{arrayMember}, \cinline{listDelEnd}, \cinline{listDelBeg},
    \cinline{addLast}, \cinline{bstSkewed},
    \cinline{popHighest}. }
  \label{fig:evaluation-nostrict}
  \renewcommand{\arraystretch}{1.0}
   \begin{adjustbox} {max width=\columnwidth,keepaspectratio}
   \begin{tabular}{|l|c|c|c|c|c|c|c|r}
    \hline
     \textbf{Framework} & \textbf{arMem} & \textbf{addL} & \textbf{bstS} & \textbf{lDelE} & \textbf{lDelB} & \textbf{popH} \\
     \hline
     GP-SA-H                 & \NoOK     & \OK       & \NoOK  & \OK       & \NoOK   & \OK    \\
     GP+SA-H                 & \OK       & \NoOK     & \NoOK  & \OK       & \OK      & \OK    \\
     \GenProof{}             & \NoOK     & \NoOK     & \OK    & \OK    & \OK         & \OK  \\
     \hline \hline
     SA++                    & \OK       & \OK      & \NoOK     & \NoOK    & \NoOK  & \NoOK  \\
     \hline
  \end{tabular}
  \end{adjustbox}
  \vspace{-0.5em}
\end{table}

\autoref{fig:evaluation-nostrict} presents the results of each
approach for the 6 benchmarks that violate our total ordering. The
verification failures on these benchmarks can be divided into two
categories. First, some approaches failed to correctly instantiate an
existential variable, as discussed in \autoref{sec:RQ1}
(\cinline{arrayMember}, \cinline{addLast},
\cinline{bstSkewed}). Second, the remaining failures were due to an
inability to identify and apply the helper lemma
(\cinline{listDelBeg}, \cinline{listDelEnd} and
\cinline{popHighest}). The failures of SA++ are due to its reliance on
a fixed set of heuristics to perform both of these tasks --- in the
case of \cinline{popHighest}, for example, the tactic did not include
a necessary helper lemma in its built-in database of auxiliary
facts. The LLM-based approaches, in contrast, were able to identify
the lemma needed to completely verify this program. On the other hand,
these approaches were less effective at supplying the right
existential witnesses in some cases. Interestingly, in some cases
the simpler approaches were able to find witnesses that the more
powerful \GenProof{} could not. This is most likely due to the
nondeterministic nature of LLMs, causing \verifyLLM{} to occasionally
fail to find the right witness.

\subsection{Discussion}
\subsubsection{Performance of Generated Code}
As with most deductive synthesizers, \synver{} prioritizes
verifiability over performance when generating candidate programs, and
our experiments suggest that it achieves this goal. That said, the
performance of synthesized programs remains an important concern. To
better understand the performance of the programs generated by
\synver{}, we investigated the impact of our requirement that
candidate programs use recursion instead of loops. For the 21
specifications that admit tail recursive implementations, we found
that \synver{} consistently generated tail recursive programs. Manual
inspection of the compiled programs indicates that GCC
\cite{gough2004introduction} was able to optimize all of these into
versions that were equivalent to those produced by an implementation
using loops.  This suggests that modern compilers can be effective at
mitigating the performance impact of \synver{}'s preference for
recursive programs.


\subsubsection{Comparison with SUSLik}
SUSLik~\cite{suslik2019} is a state-of-the-art deductive synthesizer
for generating heap-manipulating C-like programs. SUSLik takes as
input the signature and separation-logic pre- and post-conditions of
the target program, written as $\{pre\}\rightsquigarrow\{post\}$, and
searches for a corresponding implementation. This search is carried
out by applying a series of deductive synthesis rules which decompose
the current synthesis task into subtasks. At each search step, the
synthesizer examines the current goal and uses heuristics to select
the next rule to apply. The space of programs that SUSLik considers is
thus constrained both by its set of synthesis rules and the heuristics
used to apply those rules. To compare these two approaches, we applied
SUSLik on our basic and heap-manipulating benchmarks--- SUSLik does
not include rules for reasoning about functions calls and thus can not
handle any program from the API benchmark.  Of these benchmarks,
SUSLik was able successfully synthesize implementations of 9 of our
basic and 22 of our heap-manipulating
benchmarks.\footnote{These results use slightly weaker specifications
  for the \cinline{insert} and \cinline{delete} programs than
  \synver{}, as they do not guarantee the order of the elements in the
  resulting list.}

To illustrate some of the differences between SUSLik and \synver{}, we
highlight two of the programs that SUSLik was not able to solve:
\paragraph{Allocate a block of n elements}
The following specification describes a function that allocates a set
of $n$ blocks. \par\nobreak %
\vspace{-.2cm}
{\footnotesize
  \[\{0 \le n \} \rightsquigarrow \{0 \le n \, ; ret \, \mapsto x \, *
    \, sll(x,n) \}\] }%
Solving this goal requires performing case analysis on $n$, but
SUSLik's rule for case splitting requires the precondition to include
at least one heap predicate. Since the precondition here only contains
pure predicates, SUSLik cannot apply its case analysis rule, and fails
to synthesize a solution.

\paragraph{List membership}
When attempting to synthesize a program that checks whether a set $s$
with $n$ elements contains the integer $mem$, SUSLik will encounter
the following subgoal: \par\nobreak %
\vspace{-.2cm} {\footnotesize
  \begin{align*}
    \{& mem = v \, \wedge \, 0 \le n1 \, \wedge \, n = 1 + n1 \, \wedge x
        \neq null \, \wedge s = v1 \, \cup \, s1; \, \\
      & ~~~~ret \mapsto a \, * \, x
        \mapsto v \, * \, x \mapsto nxt \, * \, list\_mem(nxt,mem,n1,s1)\}
    \\
    \rightsquigarrow %
    \{ & ret \mapsto (1 + n1 = 0 \, ? \, 0 \, : \, 1) \, * \,
         list\_mem(nxt,mem,n,v \, \cup \,s1)\}
  \end{align*}}%
\noindent Here, SUSLik fails to infer that the function should return
$1$, as $(1 + n1 = 0)$ must be false. SUSLik has limited support for
reasoning about the pure fragment of specifications, and tries to
heuristically instantiate the return value of the function with either
a constant drawn from a predetermined set or one of the variables in
scope, e.g., $mem$ and $v$. None of these are the correct choice, and
SUSLik fails to solve the synthesis goal.


\subsubsection{Threats to Validity}

\paragraph{Internal validity}

We cannot guarantee the absence of \textbf{data leak} - i.e.,  the training set \coderLLM{}
used by \synver{} excludes the target implementation (specifically the programs)
of our benchmark specifications. In fact, many of our data points are based on
standard and widely used data structure implementations; therefore resulting in all the
programs being generated correctly on the first try.

\synver{} uses LLMs to generate
and verify programs that are inherently \textbf{non-deterministic}. Although our
evaluation uses fixed seeds to support reproducibility, we cannot
guarantee that our results are robust to changes in those seeds. This
contrasts with traditional program synthesizers, which typically
produce deterministic results.

\paragraph{External validity}
\synver{} uses VST to verify generated programs. Thus, both \sepauto{}
and the tactics predicted by the prover LLM are geared to VST
specifically and Rocq more broadly. Alternative Rocq-based
verification frameworks for C programs also exist, e.g., Iris
\cite{jung2018iris} and RefinedC \cite{sammler2021refinedc}. Since
proofs in these frameworks are similar to those in VST, our approach
should naturally generalize to those settings, although the prompts
for the prover LLM and \sepauto{} would need to be adapted to account
for proof idioms and the custom tactics provided by those frameworks.

\subsubsection{Limitations}
\label{sec:lim}

VST has built-in support for basic types, arrays, and pointers to a
single memory location. Reasoning about data structures that reside in
non-contiguous memory locations requires manual effort to encode the
data type in separation logic and to prove helper lemmas for reasoning
about values of that type.  At present, we have manually encoded and
verified Singly Linked Lists (SLL) and Binary Search Trees (BST);
\sepauto{} is equipped with helper lemmas for discharging obligations
related to those data structures. In order to support other data
structures, \synver{} would need to be extended with the required
specifications and helper lemmas. Similar helper lemmas would be
needed to reason about composite data types, e.g., a graph data
structure that uses adjacency list implemented as an array of singly
linked lists.

\section{Related Work}
\label{sec:relwork}
\subsection{Machine Learning for Interactive Theorem Proving}

Many of the initial learning-based proof automation techniques for
interactive theorem provers focused on the problems of \textit{premise
  selection}~\cite{AHKTU+14, KG+15, WTWD+17, PMA+23}, i.e.,
identifying lemmas that are relevant to a given theorem or proof
state, and \emph{tactic prediction}, i.e., choosing the best tactic to
apply in a given proof state. Early tactic predication works explored
different neural encodings of an in-progress proof, including
GNNs~\cite{blaauwbroek2020tactician} RNNs~\cite{huang2018gamepad,
  SASL+20} and Tree-LSTMs~\cite{yang2019coqgym, tactokOOPSLA20,
  FB+22}, as well as how to enhance the proof state with additional
information, e.g., the current partial proof~\cite{tactokOOPSLA20},
the identifiers currently appearing in a goal~\cite{passport2023}, and
recent proof scripts~\cite{blaauwbroek2020tactician}.

More recent works have also explored the use of LLMs for tactic
prediction: Copra~\cite{Copra}, for example, asks GPT-4 to predict the
next tactic in a Rocq proof script, given the current proof state,
previous proof steps, and any relevant error
messages. LeanDojo~\cite{yang2023leandojo} similarly queries an LLM to
suggest the next step in a proof in Lean~\cite{de2015lean}; both
approaches attempt to generate complete proofs by combining tactic
prediction with a heuristic search that explores different proof
directions. Other LLM-based approaches attempt to generate a complete
proof by first generating a candidate proof script and then repairing
any errors reported by the proof assistant~\cite{baldur23,
  PALM+24}. Prior LLM-based approaches also leverage purely symbolic
automation~\cite{Jiang2022ThorWH, PALM+24} when generating proofs,
although these works rely on hammers~\cite{Jiang2022ThorWH,
  czajka2018hammer}, tactics that attempt to completely discharge
subgoals in the proof assistant by calling out to external automated
theorem provers. Unlike \sepauto{}, which is designed to handle
VST-specific goals, hammers are general-purpose tools meant to
discharge arbitrary proof conditions. Similarly, all of these prior
works are trained and evaluated on corpuses of generic
theorems~\cite{yang2019coqgym, zheng2021minif2f, Putnam+Bench} drawn
from a diverse set of problem domains, including pure mathematics,
programming language metatheory, and verification of pure functional
programs.

\subsection{Machine Learning for Program Verification}

Learning-based approaches for program verification have primarily
targeted frameworks that rely on explicit annotations, e.g., loop
invariants, to enable automated reasoning, e.g., Dafny~\cite{Dafny},
Frama C~\cite{kirchner2015frama}, VeriFast~\cite{jacobs2011verifast},
and Verus~\cite{lattuada2023verus}. These annotations are used to
generate \emph{verification conditions}--- formulas in a decidable
logic whose validity guarantees the correctness of the original
program--- that can be discharged by a automated theorem prover like
Z3. A key challenge when using these tools is identifying the right
set of annotations needed to verify a program. This task has
traditionally fallen to the user, but a number of tools have recently
been proposed for automatically generating these annotations.
Code2Inv~\cite{si2020code2inv}, for example, combines Graph
Representation Learning and reinforcement learning to automatically
infer loop invariants for C programs. In the past few years, several
tools have relied on LLMs to generate annotations for program written
in both C~\cite{autoSpecCAV24, veCoGen} and Rust~\cite{AutoVerus,
  safeICLR25}. A recent study suggests that LLMs struggle with
inferring specifications in full separation
logic~\cite{rego2025evaluating}, resulting in hundreds of compilation
and verification errors across different prompts. As a consequence,
most of these tools use simpler assertion languages than separation
logic, preventing them from reasoning about the full range of
specifications that \synver{} currently supports --- Frama C, for
example, cannot handle the separating implication operator, and
requires increasingly complex annotations for non-contiguous heap
allocated structures.  A verification case study~\cite{framaCContiki}
of the linked list module of Contiki~\cite{dunkels2004contiki} in
Frama C, for example, required 1400 lines of annotations to verify 176
lines of C code.

\subsection{Program Synthesis}
Like \synver{}, traditional deductive synthesizers also generate
correct-by-construction from formal specifications~\cite{KIDS,
  SpecWareGC, Spiral, fiatPOPL2015, narcissusDelaware2019, suslik2019,
  cobaltMishra22}, but unlike \synver{}, these works either target
specific application domains to achieve automation or rely on user
guidance to synthesize a program. Recent work has shown that LLMs are
effective at synthesizing programs with less rigorous specifications,
e.g., input-output examples~\cite{austin2021LLMPBE,
  testDrivenEmpirical}.  LLMs have also shown potential to
simultaneously generate programs and their specifications in
solver-aided languages from natural language
descriptions~\cite{LLMDafny2024, chakraborty2024towards}, although the
LLM-generated specifications tend to be weaker than those used by
\synver{}, however. Even when the resulting programs can be verified
against their generated specifications there is no guarantee that
those specifications accurately capture the user's intent.

\section{Conclusion}
\label{sec:conc}
We have presented \synver{}, a program synthesis framework that is
capable of generating high-assurance C programs for a diverse range of
problem domains. Each synthesized program is accompanied by a formal
proof--- built using the Rocq-based VST framework--- that ensures it
satisfies its target separation logic specification. To accomplish
this, \synver{} employs two LLMs: a coder LLM which generates
candidate programs from user-provided specifications, and a prover LLM
which helps automatically verify the correctness of candidate
programs. To facilitate verification, \synver{} places a set of
syntactic biases on generated programs that make them amenable to
automated reasoning. \synver{} verifies programs using a hybrid
strategy that combines symbolic reasoning with LLM-assisted proof
generation, causing it to discharge proof obligations that neither
approach can handle on its own. We have demonstrated the applicability
of \synver{} on a diverse set of benchmarks drawn from the program
synthesis and verification literature.

\section*{Acknowledgements}
The authors thank the anonymous reviewers for their valuable
suggestions and feedback. The authors thank Andrew Appel for his help
with VST and his insights on tail call optimization in GCC. The
authors also thank Robert Thompson for helping us set up the Rango
evaluations. This work was partially supported by the National Science
Foundation under Grant CCF-SHF 2321680 and by a grant from the Purdue
Research Foundation.

\bibliography{IEEEabrv, bibliography}

\begin{thebibliography}{10}
\providecommand{\url}[1]{#1}
\csname url@samestyle\endcsname
\providecommand{\newblock}{\relax}
\providecommand{\bibinfo}[2]{#2}
\providecommand{\BIBentrySTDinterwordspacing}{\spaceskip=0pt\relax}
\providecommand{\BIBentryALTinterwordstretchfactor}{4}
\providecommand{\BIBentryALTinterwordspacing}{\spaceskip=\fontdimen2\font plus
\BIBentryALTinterwordstretchfactor\fontdimen3\font minus \fontdimen4\font\relax}
\providecommand{\BIBforeignlanguage}[2]{{%
\expandafter\ifx\csname l@#1\endcsname\relax
\typeout{** WARNING: IEEEtran.bst: No hyphenation pattern has been}%
\typeout{** loaded for the language `#1'. Using the pattern for}%
\typeout{** the default language instead.}%
\else
\language=\csname l@#1\endcsname
\fi
#2}}
\providecommand{\BIBdecl}{\relax}
\BIBdecl

\bibitem{MW+79}
\BIBentryALTinterwordspacing
Z.~Manna and R.~Waldinger, ``Synthesis: Dreams ⇒ programs,'' \emph{IEEE Trans. Softw. Eng.}, vol.~5, no.~4, p. 294–328, Jul. 1979. [Online]. Available: \url{https://doi.org/10.1109/TSE.1979.234198}
\BIBentrySTDinterwordspacing

\bibitem{fiatPOPL2015}
\BIBentryALTinterwordspacing
B.~Delaware, C.~Pit-Claudel, J.~Gross, and A.~Chlipala, ``Fiat: Deductive synthesis of abstract data types in a proof assistant,'' \emph{SIGPLAN Not.}, vol.~50, no.~1, p. 689–700, jan 2015. [Online]. Available: \url{https://doi.org/10.1145/2775051.2677006}
\BIBentrySTDinterwordspacing

\bibitem{suslik2019}
N.~Polikarpova, ``Suslik: Synthesis of safe pointer-manipulating programs (invited tutorial),'' in \emph{2019 Formal Methods in Computer Aided Design (FMCAD)}, 2019, pp. 1--1.

\bibitem{suslikICFP21}
\BIBentryALTinterwordspacing
Y.~Watanabe, K.~Gopinathan, G.~P\^{\i}rlea, N.~Polikarpova, and I.~Sergey, ``Certifying the synthesis of heap-manipulating programs,'' \emph{Proc. ACM Program. Lang.}, vol.~5, no. ICFP, aug 2021. [Online]. Available: \url{https://doi.org/10.1145/3473589}
\BIBentrySTDinterwordspacing

\bibitem{narcissusDelaware2019}
\BIBentryALTinterwordspacing
B.~Delaware, S.~Suriyakarn, C.~Pit-Claudel, Q.~Ye, and A.~Chlipala, ``Narcissus: correct-by-construction derivation of decoders and encoders from binary formats,'' \emph{Proc. ACM Program. Lang.}, vol.~3, no. ICFP, jul 2019. [Online]. Available: \url{https://doi.org/10.1145/3341686}
\BIBentrySTDinterwordspacing

\bibitem{cobaltMishra22}
\BIBentryALTinterwordspacing
A.~Mishra and S.~Jagannathan, ``Specification-guided component-based synthesis from effectful libraries,'' \emph{Proc. ACM Program. Lang.}, vol.~6, no. OOPSLA2, oct 2022. [Online]. Available: \url{https://doi.org/10.1145/3563310}
\BIBentrySTDinterwordspacing

\bibitem{SpecWareGC}
D.~Pavlovic, P.~Pepper, and D.~R. Smith, ``Formal derivation of concurrent garbage collectors,'' in \emph{Mathematics of Program Construction}.\hskip 1em plus 0.5em minus 0.4em\relax Springer Berlin Heidelberg, 2010, pp. 353--376.

\bibitem{LLMDafny2024}
\BIBentryALTinterwordspacing
M.~R.~H. Misu, C.~V. Lopes, I.~Ma, and J.~Noble, ``Towards ai-assisted synthesis of verified dafny methods,'' \emph{Proc. {ACM} Softw. Eng.}, vol.~1, no. {FSE}, pp. 812--835, 2024. [Online]. Available: \url{https://doi.org/10.1145/3643763}
\BIBentrySTDinterwordspacing

\bibitem{chakraborty2024towards}
\BIBentryALTinterwordspacing
S.~Chakraborty, G.~Ebner, S.~Bhat, S.~Fakhoury, S.~Fatima, S.~K. Lahiri, and N.~Swamy, ``Towards neural synthesis for smt-assisted proof-oriented programming,'' in \emph{47th {IEEE/ACM} International Conference on Software Engineering, {ICSE} 2025, Ottawa, ON, Canada, April 26 - May 6, 2025}.\hskip 1em plus 0.5em minus 0.4em\relax {IEEE}, 2025, pp. 1755--1767. [Online]. Available: \url{https://doi.org/10.1109/ICSE55347.2025.00002}
\BIBentrySTDinterwordspacing

\bibitem{AutoVerus}
\BIBentryALTinterwordspacing
C.~Yang, X.~Li, M.~R.~H. Misu, J.~Yao, W.~Cui, Y.~Gong, C.~Hawblitzel, S.~K. Lahiri, J.~R. Lorch, S.~Lu, F.~Yang, Z.~Zhou, and S.~Lu, ``Autoverus: Automated proof generation for rust code,'' \emph{CoRR}, vol. abs/2409.13082, 2024. [Online]. Available: \url{https://doi.org/10.48550/arXiv.2409.13082}
\BIBentrySTDinterwordspacing

\bibitem{veCoGen}
\BIBentryALTinterwordspacing
M.~Sevenhuijsen, K.~Etemadi, and M.~Nyberg, ``Vecogen: Automating generation of formally verified c code with large language models,'' 2025. [Online]. Available: \url{https://arxiv.org/abs/2411.19275}
\BIBentrySTDinterwordspacing

\bibitem{autoSpecCAV24}
C.~Wen, J.~Cao, J.~Su, Z.~Xu, S.~Qin, M.~He, H.~Li, S.-C. Cheung, and C.~Tian, ``Enchanting program specification synthesis by large language models using static analysis and program verification,'' in \emph{Computer Aided Verification}, A.~Gurfinkel and V.~Ganesh, Eds.\hskip 1em plus 0.5em minus 0.4em\relax Cham: Springer Nature Switzerland, 2024, pp. 302--328.

\bibitem{reynolds2002}
J.~Reynolds, ``Separation logic: a logic for shared mutable data structures,'' in \emph{Proceedings 17th Annual IEEE Symposium on Logic in Computer Science}, 2002, pp. 55--74.

\bibitem{vst2018appel}
\BIBentryALTinterwordspacing
Q.~Cao, L.~Beringer, S.~Gruetter, J.~Dodds, and A.~W. Appel, ``Vst-floyd: {A} separation logic tool to verify correctness of {C} programs,'' \emph{J. Autom. Reason.}, vol.~61, no. 1-4, pp. 367--422, 2018. [Online]. Available: \url{https://doi.org/10.1007/s10817-018-9457-5}
\BIBentrySTDinterwordspacing

\bibitem{Coq-refman}
{The Coq Development Team}, ``The {Coq} reference manual -- release 8.19.0,'' \url{https://coq.inria.fr/doc/V8.19.0/refman}, 2024.

\bibitem{yang2023leandojo}
K.~Yang, A.~Swope, A.~Gu, R.~Chalamala, P.~Song, S.~Yu, S.~Godil, R.~Prenger, and A.~Anandkumar, ``{LeanDojo}: Theorem proving with retrieval-augmented language models,'' in \emph{Neural Information Processing Systems (NeurIPS)}, 2023.

\bibitem{baldur23}
\BIBentryALTinterwordspacing
E.~First, M.~N. Rabe, T.~Ringer, and Y.~Brun, ``Baldur: Whole-proof generation and repair with large language models,'' in \emph{Proceedings of the 31st {ACM} Joint European Software Engineering Conference and Symposium on the Foundations of Software Engineering, {ESEC/FSE} 2023, San Francisco, CA, USA, December 3-9, 2023}, S.~Chandra, K.~Blincoe, and P.~Tonella, Eds.\hskip 1em plus 0.5em minus 0.4em\relax {ACM}, 2023, pp. 1229--1241. [Online]. Available: \url{https://doi.org/10.1145/3611643.3616243}
\BIBentrySTDinterwordspacing

\bibitem{synver:artifact}
\BIBentryALTinterwordspacing
P.~Mukherjee, M.~Lu, and B.~Delaware, ``{LLM-Assisted Synthesis of High-Assurance C Programs},'' sep 2025. [Online]. Available: \url{https://doi.org/10.5281/zenodo.17219749}
\BIBentrySTDinterwordspacing

\bibitem{hoare1969}
C.~A.~R. Hoare, ``An axiomatic basis for computer programming,'' \emph{Commun. ACM}, vol.~12, no.~10, p. 576–580, Oct. 1969.

\bibitem{cao2018vst}
Q.~Cao, L.~Beringer, S.~Gruetter, J.~Dodds, and A.~W. Appel, ``Vst-floyd: A separation logic tool to verify correctness of c programs,'' \emph{Journal of Automated Reasoning}, vol.~61, pp. 367--422, 2018.

\bibitem{furia2014loop}
C.~A. Furia, B.~Meyer, and S.~Velder, ``Loop invariants: Analysis, classification, and examples,'' \emph{ACM Computing Surveys (CSUR)}, vol.~46, no.~3, pp. 1--51, 2014.

\bibitem{xiao2023hyperwand}
Y.~Xiao, ``Hyperwand: Extending the magic wand operator in separation logic,'' \emph{j}, 2023.

\bibitem{rangoICSE2025}
\BIBentryALTinterwordspacing
K.~Thompson, N.~Saavedra, P.~Carrott, K.~Fisher, A.~Sanchez{-}Stern, Y.~Brun, J.~F. Ferreira, S.~Lerner, and E.~First, ``Rango: Adaptive retrieval-augmented proving for automated software verification,'' \emph{ICSE}, 2025. [Online]. Available: \url{https://people.cs.umass.edu/~brun/pubs/pubs/Thompson25icse.pdf}
\BIBentrySTDinterwordspacing

\bibitem{clrs}
T.~H. Cormen, C.~E. Leiserson, R.~L. Rivest, and C.~Stein, \emph{Introduction to Algorithms, Third Edition}, 3rd~ed.\hskip 1em plus 0.5em minus 0.4em\relax The MIT Press, 2009.

\bibitem{blaauwbroek2020tactician}
L.~Blaauwbroek, J.~Urban, and H.~Geuvers, ``The tactician: A seamless, interactive tactic learner and prover for coq,'' in \emph{International Conference on Intelligent Computer Mathematics}.\hskip 1em plus 0.5em minus 0.4em\relax Springer, 2020, pp. 271--277.

\bibitem{CPDT}
A.~Chlipala, \emph{Certified Programming with Dependent Types: A Pragmatic Introduction to the Coq Proof Assistant}.\hskip 1em plus 0.5em minus 0.4em\relax The MIT Press, 2013.

\bibitem{gough2004introduction}
B.~J. Gough and R.~Stallman, \emph{An Introduction to GCC.}\hskip 1em plus 0.5em minus 0.4em\relax Network Theory Limited Bristol, UK, 2004.

\bibitem{jung2018iris}
R.~Jung, R.~Krebbers, J.-H. Jourdan, A.~Bizjak, L.~Birkedal, and D.~Dreyer, ``Iris from the ground up: A modular foundation for higher-order concurrent separation logic,'' \emph{Journal of Functional Programming}, vol.~28, p. e20, 2018.

\bibitem{sammler2021refinedc}
M.~Sammler, R.~Lepigre, R.~Krebbers, K.~Memarian, D.~Dreyer, and D.~Garg, ``Refinedc: automating the foundational verification of c code with refined ownership types,'' in \emph{Proceedings of the 42nd ACM SIGPLAN International Conference on Programming Language Design and Implementation}, 2021, pp. 158--174.

\bibitem{AHKTU+14}
\BIBentryALTinterwordspacing
J.~Alama, T.~Heskes, D.~K\"{u}hlwein, E.~Tsivtsivadze, and J.~Urban, ``Premise selection for mathematics by corpus analysis and kernel methods,'' \emph{J. Autom. Reason.}, vol.~52, no.~2, p. 191–213, Feb. 2014. [Online]. Available: \url{https://doi.org/10.1007/s10817-013-9286-5}
\BIBentrySTDinterwordspacing

\bibitem{KG+15}
\BIBentryALTinterwordspacing
T.~Gauthier and C.~Kaliszyk, ``Premise selection and external provers for hol4,'' in \emph{Proceedings of the 2015 Conference on Certified Programs and Proofs}, ser. CPP '15.\hskip 1em plus 0.5em minus 0.4em\relax New York, NY, USA: Association for Computing Machinery, 2015, p. 49–57. [Online]. Available: \url{https://doi.org/10.1145/2676724.2693173}
\BIBentrySTDinterwordspacing

\bibitem{WTWD+17}
M.~Wang, Y.~Tang, J.~Wang, and J.~Deng, ``Premise selection for theorem proving by deep graph embedding,'' in \emph{Proceedings of the 31st International Conference on Neural Information Processing Systems}, ser. NIPS'17.\hskip 1em plus 0.5em minus 0.4em\relax Red Hook, NY, USA: Curran Associates Inc., 2017, p. 2783–2793.

\bibitem{PMA+23}
\BIBentryALTinterwordspacing
B.~Piotrowski, R.~F. Mir, and E.~Ayers, ``Machine-learned premise selection for lean,'' in \emph{Automated Reasoning with Analytic Tableaux and Related Methods: 32nd International Conference, TABLEAUX 2023, Prague, Czech Republic, September 18–21, 2023, Proceedings}.\hskip 1em plus 0.5em minus 0.4em\relax Berlin, Heidelberg: Springer-Verlag, 2023, p. 175–186. [Online]. Available: \url{https://doi.org/10.1007/978-3-031-43513-3_10}
\BIBentrySTDinterwordspacing

\bibitem{huang2018gamepad}
D.~Huang, P.~Dhariwal, D.~Song, and I.~Sutskever, ``Gamepad: A learning environment for theorem proving,'' \emph{arXiv preprint arXiv:1806.00608}, 2018.

\bibitem{SASL+20}
\BIBentryALTinterwordspacing
A.~Sanchez-Stern, Y.~Alhessi, L.~Saul, and S.~Lerner, ``Generating correctness proofs with neural networks,'' in \emph{Proceedings of the 4th ACM SIGPLAN International Workshop on Machine Learning and Programming Languages}, ser. MAPL 2020.\hskip 1em plus 0.5em minus 0.4em\relax New York, NY, USA: Association for Computing Machinery, 2020, p. 1–10. [Online]. Available: \url{https://doi.org/10.1145/3394450.3397466}
\BIBentrySTDinterwordspacing

\bibitem{yang2019coqgym}
K.~Yang and J.~Deng, ``Learning to prove theorems via interacting with proof assistants,'' in \emph{International Conference on Machine Learning (ICML)}, 2019.

\bibitem{tactokOOPSLA20}
\BIBentryALTinterwordspacing
E.~First, Y.~Brun, and A.~Guha, ``Tactok: semantics-aware proof synthesis,'' \emph{Proc. {ACM} Program. Lang.}, vol.~4, no. {OOPSLA}, pp. 231:1--231:31, 2020. [Online]. Available: \url{https://doi.org/10.1145/3428299}
\BIBentrySTDinterwordspacing

\bibitem{FB+22}
\BIBentryALTinterwordspacing
E.~First and Y.~Brun, ``Diversity-driven automated formal verification,'' in \emph{Proceedings of the 44th International Conference on Software Engineering}, ser. ICSE '22.\hskip 1em plus 0.5em minus 0.4em\relax New York, NY, USA: Association for Computing Machinery, 2022, p. 749–761. [Online]. Available: \url{https://doi.org/10.1145/3510003.3510138}
\BIBentrySTDinterwordspacing

\bibitem{passport2023}
\BIBentryALTinterwordspacing
A.~Sanchez{-}Stern, E.~First, T.~Zhou, Z.~Kaufman, Y.~Brun, and T.~Ringer, ``Passport: Improving automated formal verification using identifiers,'' \emph{{ACM} Trans. Program. Lang. Syst.}, vol.~45, no.~2, pp. 12:1--12:30, 2023. [Online]. Available: \url{https://doi.org/10.1145/3593374}
\BIBentrySTDinterwordspacing

\bibitem{Copra}
\BIBentryALTinterwordspacing
A.~Thakur, G.~Tsoukalas, Y.~Wen, J.~Xin, and S.~Chaudhuri, ``An in-context learning agent for formal theorem-proving,'' 2024. [Online]. Available: \url{https://arxiv.org/abs/2310.04353}
\BIBentrySTDinterwordspacing

\bibitem{de2015lean}
L.~De~Moura, S.~Kong, J.~Avigad, F.~Van~Doorn, and J.~von Raumer, ``{The Lean theorem prover (system description)},'' in \emph{Automated Deduction-CADE-25: 25th International Conference on Automated Deduction, Berlin, Germany, August 1-7, 2015, Proceedings 25}.\hskip 1em plus 0.5em minus 0.4em\relax Springer, 2015, pp. 378--388.

\bibitem{PALM+24}
\BIBentryALTinterwordspacing
M.~Lu, B.~Delaware, and T.~Zhang, ``Proof automation with large language models,'' in \emph{Proceedings of the 39th IEEE/ACM International Conference on Automated Software Engineering}, ser. ASE '24.\hskip 1em plus 0.5em minus 0.4em\relax New York, NY, USA: Association for Computing Machinery, 2024, p. 1509–1520. [Online]. Available: \url{https://doi.org/10.1145/3691620.3695521}
\BIBentrySTDinterwordspacing

\bibitem{Jiang2022ThorWH}
A.~Jiang, K.~Czechowski, M.~Jamnik, P.~Milos, S.~Tworkowski, W.~Li, and Y.~T. Wu, ``Thor: Wielding hammers to integrate language models and automated theorem provers,'' in \emph{NeurIPS}, 2022.

\bibitem{czajka2018hammer}
\BIBentryALTinterwordspacing
z.~Czajka and C.~Kaliszyk, ``Hammer for coq: Automation for dependent type theory,'' \emph{J. Autom. Reason.}, vol.~61, no. 1–4, p. 423–453, Jun. 2018. [Online]. Available: \url{https://doi.org/10.1007/s10817-018-9458-4}
\BIBentrySTDinterwordspacing

\bibitem{zheng2021minif2f}
K.~Zheng, J.~M. Han, and S.~Polu, ``Minif2f: a cross-system benchmark for formal olympiad-level mathematics,'' \emph{arXiv preprint arXiv:2109.00110}, 2021.

\bibitem{Putnam+Bench}
G.~Tsoukalas, J.~Lee, J.~Jennings, J.~Xin, M.~Ding, M.~Jennings, A.~Thakur, and S.~Chaudhuri, ``Putnambench: evaluating neural theorem-provers on the putnam mathematical competition,'' in \emph{Proceedings of the 38th International Conference on Neural Information Processing Systems}, ser. NIPS '24.\hskip 1em plus 0.5em minus 0.4em\relax Red Hook, NY, USA: Curran Associates Inc., 2025.

\bibitem{Dafny}
K.~R.~M. Leino, ``Dafny: an automatic program verifier for functional correctness,'' in \emph{Proceedings of the 16th International Conference on Logic for Programming, Artificial Intelligence, and Reasoning}, ser. LPAR'10.\hskip 1em plus 0.5em minus 0.4em\relax Berlin, Heidelberg: Springer-Verlag, 2010, p. 348–370.

\bibitem{kirchner2015frama}
F.~Kirchner, N.~Kosmatov, V.~Prevosto, J.~Signoles, and B.~Yakobowski, ``Frama-c: A software analysis perspective,'' \emph{Formal aspects of computing}, vol.~27, no.~3, pp. 573--609, 2015.

\bibitem{jacobs2011verifast}
B.~Jacobs, J.~Smans, P.~Philippaerts, F.~Vogels, W.~Penninckx, and F.~Piessens, ``Verifast: A powerful, sound, predictable, fast verifier for c and java,'' in \emph{NASA formal methods symposium}.\hskip 1em plus 0.5em minus 0.4em\relax Springer, 2011, pp. 41--55.

\bibitem{lattuada2023verus}
A.~Lattuada, T.~Hance, C.~Cho, M.~Brun, I.~Subasinghe, Y.~Zhou, J.~Howell, B.~Parno, and C.~Hawblitzel, ``Verus: Verifying rust programs using linear ghost types,'' \emph{Proceedings of the ACM on Programming Languages}, vol.~7, no. OOPSLA1, pp. 286--315, 2023.

\bibitem{si2020code2inv}
X.~Si, A.~Naik, H.~Dai, M.~Naik, and L.~Song, ``Code2inv: A deep learning framework for program verification,'' in \emph{Computer Aided Verification: 32nd International Conference, CAV 2020, Los Angeles, CA, USA, July 21--24, 2020, Proceedings, Part II 32}.\hskip 1em plus 0.5em minus 0.4em\relax Springer, 2020, pp. 151--164.

\bibitem{safeICLR25}
\BIBentryALTinterwordspacing
T.~Chen, S.~Lu, S.~Lu, Y.~Gong, C.~Yang, X.~Li, M.~R.~H. Misu, H.~Yu, N.~Duan, P.~Cheng, F.~Yang, S.~K. Lahiri, T.~Xie, and L.~Zhou, ``Automated proof generation for rust code via self-evolution,'' in \emph{The Thirteenth International Conference on Learning Representations, {ICLR} 2025, Singapore, April 24-28, 2025}.\hskip 1em plus 0.5em minus 0.4em\relax OpenReview.net, 2025. [Online]. Available: \url{https://openreview.net/forum?id=2NqssmiXLu}
\BIBentrySTDinterwordspacing

\bibitem{rego2025evaluating}
M.~Rego, W.~Fan, X.~Hu, S.~Dod, Z.~Ni, D.~Xie, J.~DiVincenzo, and L.~Tan, ``Evaluating the ability of gpt-4o to generate verifiable specifications in verifast,'' in \emph{2025 IEEE/ACM Second International Conference on AI Foundation Models and Software Engineering (Forge)}.\hskip 1em plus 0.5em minus 0.4em\relax IEEE, 2025, pp. 246--251.

\bibitem{framaCContiki}
A.~Blanchard, N.~Kosmatov, and F.~Loulergue, ``Ghosts for lists: A critical module of contiki verified in frama-c,'' in \emph{NASA Formal Methods}, A.~Dutle, C.~Mu{\~{n}}oz, and A.~Narkawicz, Eds.\hskip 1em plus 0.5em minus 0.4em\relax Cham: Springer International Publishing, 2018, pp. 37--53.

\bibitem{dunkels2004contiki}
A.~Dunkels, B.~Gronvall, and T.~Voigt, ``Contiki-a lightweight and flexible operating system for tiny networked sensors,'' in \emph{29th annual IEEE international conference on local computer networks}.\hskip 1em plus 0.5em minus 0.4em\relax IEEE, 2004, pp. 455--462.

\bibitem{KIDS}
D.~Smith, ``Kids: a semiautomatic program development system,'' \emph{IEEE Transactions on Software Engineering}, vol.~16, no.~9, pp. 1024--1043, 1990.

\bibitem{Spiral}
F.~Franchetti, T.-M. Low, T.~Popovici, R.~Veras, D.~G. Spampinato, J.~Johnson, M.~P{\"u}schel, J.~C. Hoe, and J.~M.~F. Moura, ``{SPIRAL}: Extreme performance portability,'' \emph{Proceedings of the IEEE, special issue on ``From High Level Specification to High Performance Code''}, vol. 106, no.~11, 2018.

\bibitem{austin2021LLMPBE}
J.~Austin, A.~Odena, M.~Nye, M.~Bosma, H.~Michalewski, D.~Dohan, E.~Jiang, C.~Cai, M.~Terry, Q.~Le, and C.~Sutton, ``Program synthesis with large language models,'' 2021.

\bibitem{testDrivenEmpirical}
\BIBentryALTinterwordspacing
S.~Fakhoury, A.~Naik, G.~Sakkas, S.~Chakraborty, and S.~K. Lahiri, ``Llm-based test-driven interactive code generation: User study and empirical evaluation,'' \emph{{IEEE} Trans. Software Eng.}, vol.~50, no.~9, pp. 2254--2268, 2024. [Online]. Available: \url{https://doi.org/10.1109/TSE.2024.3428972}
\BIBentrySTDinterwordspacing

\end{thebibliography}


\end{document}